\documentclass[twocolumn,twocolappendix]{aastex7}
\usepackage{orcidlink}
\usepackage{graphicx}	
\usepackage{amsmath}	
\usepackage{natbib}
\usepackage{booktabs}
\usepackage{cleveref}
\usepackage{siunitx}

\newcommand{\teff}[1]{$T_{\text{eff}}$#1}
\newcommand{\prot}[1]{$P_{\text{rot}}$#1}

\newcommand{\logg}[1]{$\log{g}$#1}
\newcommand{\vsini}[1]{$v\sin{i_\star}$#1}

\newcommand{\degree}{$^{\circ}$}

\defcitealias{Canas_2022}{C22}
\defcitealias{Canas_2023}{C23}

\begin{document}

\title{Aligned Stellar Obliquities for Two Hot Jupiter-hosting M Dwarfs Revealed by MAROON-X: Implications for Hot Jupiter Formation}

\author[0000-0002-7992-469X]{Drew Weisserman}
\altaffiliation{These authors contributed equally to this work.}
\affiliation{Department of Physics \& Astronomy, McMaster University, 1280 Main St W, Hamilton, ON, L8S 4L8, Canada}
\email[show]{weisserd@mcmaster.ca}

\author[0009-0000-4040-6628]{Erik Gillis}
\altaffiliation{These authors contributed equally to this work.}
\affiliation{Department of Physics \& Astronomy, McMaster University, 1280 Main St W, Hamilton, ON, L8S 4L8, Canada}
\email[show]{gillie1@mcmaster.ca}

\author[0000-0001-5383-9393]{Ryan Cloutier}
\affiliation{Department of Physics \& Astronomy, McMaster University, 1280 Main St W, Hamilton, ON, L8S 4L8, Canada}
\email{test}

\author[0009-0003-1142-292X]{Nina Brown}
\affiliation{Department of Astronomy \& Astrophysics, University of Chicago, Chicago, IL 60637, USA}
\email{test}

\author[0000-0003-4733-6532]{Jacob L.\ Bean}
\affiliation{Department of Astronomy \& Astrophysics, University of Chicago, Chicago, IL 60637, USA}
\email{test}

\author[0000-0003-4526-3747]{Andreas Seifahrt}
\affiliation{Gemini Observatory/NSF NOIRLab, 670 N. A'ohoku Place, Hilo, HI 96720, USA}
\email{test}

\author[0009-0005-1486-8374]{Tanya Das}
\affiliation{Department of Astronomy \& Astrophysics, University of Chicago, Chicago, IL 60637, USA}
\email{test}

\author[0000-0003-2404-2427]{Madison Brady}
\affiliation{Department of Astronomy \& Astrophysics, University of Chicago, Chicago, IL 60637, USA}
\email{mtbrady@uchicago.edu}

\author[0000-0002-8868-7649]{Bertram Bitsch}
\affiliation{Department of Physics, University College Cork, College Rd, Cork T12 K8AF, Ireland}
\email{test}

\author[0000-0001-9796-2158]{Emily Deibert}
\affiliation{Gemini Observatory/NSF NOIRLab, Casilla 603, La Serena, Chile}
\email{test}

\author[0000-0001-5442-1300]{Thomas M. Evans-Soma}
\affiliation{School of Information and Physical Sciences, University of Newcastle, Callaghan, NSW, Australia}
\email{test}

\author{Noah Fenlon}
\affiliation{Department of Physics \& Astronomy, McMaster University, 1280 Main St W, Hamilton, ON, L8S 4L8, Canada}
\email{test}

\author[0000-0003-0514-1147]{Laura Kreidberg}
\affiliation{Max Planck Institute for Astronomy, Heidelberg, Germany}
\email{test}

\author[0000-0002-2338-476X]{Michael Line}
\affiliation{School of Earth and Space Exploration, Arizona State University, Tempe, AZ 85281, USA}
\email{test}

\author[0000-0002-7605-2961]{Ralph Pudritz}
\affiliation{Department of Physics \& Astronomy, McMaster University, 1280 Main St W, Hamilton, ON, L8S 4L8, Canada}
\email{test}

\author[0000-0002-7260-5821]{Evgenya L. Shkolnik}
\affiliation{School of Earth and Space Exploration, Arizona State University, Tempe, AZ 85281, USA}
\email{test}

\author[0000-0003-0156-4564]{Luis Welbanks}
\affiliation{School of Earth and Space Exploration, Arizona State University, Tempe, AZ 85281, USA}
\email{test}

\begin{abstract}
Hot Jupiters (HJs) are $2-3\times$ less common around early M dwarfs than around AFGK stars, suggesting that HJs may form and/or migrate via distinct pathways around different types of stars. One source of insight into HJ formation mechanisms is to trace their dynamical histories through measurements of host stellar obliquities via the Rossiter-McLaughlin (RM) effect. Here we present measurements of the RM effect for the HJs TOI-3714 b and TOI-5293 A b using the Gemini-North/MAROON-X spectrograph. Our measurements represent just the second and third hot Jupiters around M dwarfs (HJMD) with a detection of the RM effect. We find that both systems are well-aligned with sky-projected obliquities of $\lambda = 21^{+14}_{-11} \si{\degree}$ and $-12^{+19}_{-14} \si{\degree}$ and deprojected obliquities of $\psi = 26^{+11}_{-10} \si{\degree}$ and $24^{+11}_{-10} \si{\degree}$ for TOI-3714 and TOI-5293 A, respectively. Both stars are in wide binary systems. We refine the stellar parameters by decontaminating their unresolved $K_s$-band photometry and constrain the binary orbits using Gaia DR3 astrometry. We find that the minimum mutual inclination of the planet and binary companion in the TOI-5293 system is sufficiently large to drive Kozai-Lidov (KL) migration while the result for TOI-3714 is inconclusive. We present a population-level analysis of HJs around AFGK versus early M dwarfs and argue that KL migration is more efficient around the latter, which is expected to produce misaligned stellar obliquities in HJMD systems in the absence of efficient tidal damping. The emerging population of well-aligned HJMD hosts supports the expectation that M dwarfs, with their deep convective envelopes, do efficiently dampen misaligned obliquities. 
\end{abstract}


\section{Introduction} \label{sect:intro}
With orbital periods of $\lesssim 10$ days, hot Jupiters (HJs) present a stark contrast to the gas giants in the solar system and have continued to challenge theories of planet formation since their inception \citep{Dawson_2018}. Namely, in-situ formation of HJs is widely regarded as intractable due to the need for an enhancement of two orders of magnitude in the local surface density of solids needed to form massive cores that can trigger runaway gas accretion \citep[e.g.][]{Rafikov_2006}. While circumventing arguments that support in-situ formation have been presented \citep{Batygin_2016,Boley_2016}, ex-situ formation plus inward migration presents a compelling alternative theory. Migration may be disk-driven and dynamically cold as planets’ orbital eccentricities and obliquities (i.e. the angle between the stellar rotation axis and the planet's orbital plane normal) are strongly damped by the disk gas \citep[e.g][]{Lin_1986,Lin_1996}, or migration may be dynamically hot, as is the case with high eccentricity migration (HEM) such as planet-planet scattering \citep[e.g.][]{Rasio_1996,Weidenschilling_1996,Beauge_2012} and Kozai-Lidov migration \citep[e.g.][]{Wu_2003,Fabrycky_2007,Wu_2007}. HEM processes shape the orbital architectures of planetary systems by boosting planets’ orbital eccentricities, mutual inclinations, and obliquities such that measuring of these quantities can serve as a powerful diagnostic of a planet's dynamical history.

Mounting empirical evidence is emerging for HEM plus tidal damping as the probable migration mechanism for the formation of HJs around AFGK stars. Lines of observational evidence include the dearth of nearby planets in systems containing a HJ, which suggests that dynamical disruption through migration is likely \citep{Latham_2011,Bryan_2016,Huang_2016}, the positive correlation between HJ occurrence and widely separated massive planetary companions that are capable of exciting HEM \citep{Knutson_2014,Bryan_2016,Zink_2023}, the preferentially misaligned mutual inclinations of wide binary companions and HJs \citep{Behmard_2022,Christian_2025}, and the statistical reconstruction of the HJ eccentricity--obliquity--\teff{} distribution that is consistent with HEM plus tidal damping without the need to invoke other formation/migration processes \citep{Rice_2022}. However, HJs around M dwarfs (HJMDs)\footnote{HJMDs are also sometimes referred to in the literature as close-in Giant Exoplanets around M Stars, or GEMS. We note that although HJMDs have similarly short orbital periods to their contemporaries around AFGK stars, HJMDs may not necessarily be considered `hot' as they have typical equilibrium temperatures $T_{\mathrm{eq}} \lesssim 850$ K due to the low luminosities of their host stars.} remain largely missing from this picture. 

HJMDs are $\sim 2-3$ times less common than around AFGK stars \citep{Johnson_2010,Beleznay_2022,Bryant_2023,Gan_2023}, which suggests that HJMDs may form and/or migrate via a distinct pathway. This is corroborated by theoretical models of gas giant formation via core accretion that struggle to produce HJMDs due to the low disk masses around M stars \citep{Laughlin_2004,Morales_2019}. Investigating the dynamical histories of HJMDs and their comparison to HJs around early-type stars remains an open problem in large part due to the lack of HJMDs that are well-suited to detailed follow-up. Fortunately, NASA's Transiting Exoplanet Survey Satellite \cite[TESS;][]{Ricker_2015} has recently discovered a handful of HJMDs around nearby stars, many of which are amenable to follow-up observations to trace their formation and dynamical histories (i.e. TIC 46432937 b \& TOI-762 A b; \citealt{Hartman_2024}, TOI-3235 b; \citealt{Hobson_2023}, TOI-3714 b; \citealt{Canas_2022}, TOI-3984 A b \& TOI-5293 A b; \citealt{Canas_2023}, TOI-4201 b; \citealt{Gan_2024}, TOI-4860 b; \citealt{Almenara_2024}, TOI-519 b; \citealt{Kagetani_2023}, TOI-5205 b; \citealt{Kanodia_2023}, TOI-530 b; \citealt{Gan_2022}, TOI-5344 b; \citealt{Hartman_2024}, TOI-5634 b \& TOI-6034 b; \citealt{Kanodia_2024}, TOI-5688 A b; \citealt{Reji_2025}, TOI-6383 A b; \citealt{Bernabo_2024}). These planets come from querying the NASA Exoplanet Archive for TESS-discovered HJMDs with $P\leq 10$ days, $R_p\geq 0.7\, R_{\mathrm{Jup}}$, $M_p\leq 13\, M_{\mathrm{Jup}}$, $T_{\mathrm{eff}}\leq 3900$ K, $M_\star\leq 0.61\, M_\odot$.

The Rossiter-McLaughlin effect (RM) is a phenomenon exhibited during a spectroscopically observed planetary transit that is sensitive to the sky-projected stellar obliquity \citep{Rossiter_1924,McLaughlin_1924,Triaud_2018}. During a planetary transit, the planet occults different portions of the differentially Doppler-shifted stellar disk producing an anomalous radial velocity (RV) signal. The shape of the RM signature is sensitive to the sky-projected spin-orbit angle between the planet's orbital plane normal and the host star's rotational axis (a.k.a. the stellar obliquity). To date, the RM effect has been used to measure stellar obliquities and characterize the dynamical histories of $\sim 240$ planets\footnote{Based on the \href{https://www.astro.keele.ac.uk/jkt/tepcat/obliquity.html}{TEPCAT catalog}.} over a wide range of planetary masses. Within this sample of planetary systems with obliquity measurements, only eleven planets orbit an M dwarf with \teff{} $< 3900$ K (i.e. AU Mic b; \citealt{Hirano_2020a,Martioli_2020,Palle_2020,Addison_2021}, GJ 436 b; \citealt{Bourrier_2018,Bourrier_2022}, GJ 3470 b; \citealt{Stefansson_2022}, K2-25 b; \citealt{Gaidos_2020,Stefansson_2020b}, K2-33 b; \citealt{Hirano_2024}, TOI-3884 b; \citealt{LibbyRoberts_2023}, TRAPPIST-1 b,c,e,f; \citealt{Hirano_2020b,Brady_2023}, TOI-4201 b; \citealt{Gan_2024}), most of which have sizes within the terrestrial to Neptunian regime ($R_p< 5\, R_\oplus$). The one exception is TOI-4201, which represents the first detection of the RM effect of a HJMD.

Here we present detections of the RM effect for the HJMDs TOI-3714 b and TOI-5293 A b. Our measurements bring the number of RM detections of HJMDs from one to three. In Section~\ref{sect:stars} we present refined stellar and planetary parameters for both planetary systems. In Section~\ref{sect:data} we describe our observations with the MAROON-X spectrograph. In Section~\ref{sect:analysis} we describe our data analysis methodology. We present our results and a discussion in Section~\ref{sect:discussion} before concluding with a summary of our main findings in Section~\ref{sect:summary}.

\section{Targets} \label{sect:stars}
\subsection{The S-Type Planetary System around TOI-3714} \label{sect:system3714}
\citealt{Canas_2022} (hereafter \citetalias{Canas_2022}) reported the discovery of a hot Jupiter on an S-type orbit around the M2 dwarf companion star in the TOI-3714 binary system. S-type orbits are a class of orbital configurations in which a planet orbits one member of a binary star system, as opposed to a circumbinary orbit (i.e. P-type). The primary star in the TOI-3714 system is a white dwarf with a mass of $\approx 1.07\, M_\odot$ and an estimated cooling age of 2.4 Gyrs \citepalias{Canas_2022}. Both stellar components are resolved in Gaia DR3 with an angular separation of 3\farcs37 and a projected separation of 381 au \citep{GaiaDR3}. The system is located at a distance of $113.1\pm 0.3$ pc.

\citetalias{Canas_2022} report a model-dependent stellar mass and radius for TOI-3714 based on fits of the stellar spectral energy distribution (SED) to MIST isochrones \citep[Modules for Experiments in Stellar Astrophysics Isochrones and Stellar Tracks;][]{Choi_2016}. We elect to rederive the stellar mass and radius because model-dependent M dwarf masses and radii are known to suffer from inconsistencies with model-independent measurements \citep{Mann_2019}. We adopt an alternative method that relies on empirically calibrated and model-independent mass-$M_{K_{s}}$ and radius-$M_{K_{s}}$ relations for M dwarfs \citep{Mann_2015,Mann_2019}. Evaluating the mass and radius of TOI-3714 requires accurate measurements of the star's absolute $K_s$-band magnitude after correcting for any dilution by neighbouring stars. Querying the 2MASS point source catalog \citep{cutri03} reveals that only the M dwarf companion TOI-3714 is detected and with no photometric quality flags raised. This implies that the unresolved white dwarf primary is contaminating the 2MASS $K_s$-band magnitude of TOI-3714. 

We estimate the level of dilution of TOI-3714 by the white dwarf primary in the $K_s$-band using both stars' Gaia magntiudes ($\Delta G = 4.56$) and Gaia-2MASS color transformations \citep{Evans_2018}. Use the $(G-K_s)-(BP-RP)$ color relation and propagating errors, which includes a systematic uncertainty of 0.083 mag in the color relation, we estimate the $K_s$-band flux ratio $F_{K_s,MD}/F_{K_s,WD}=417^{+78}_{-64}$. As such, the dilution of TOI-3714's $K_s$-band flux by the white dwarf primary is negligible. We proceed with calculating the stellar mass and radius of TOI-3714 using its $K_s$-band magnitude of $10.852\pm 0.017$, Gaia DR3 distance, and the aforementioned mass-$M_{K_{s}}$ and radius-$M_{K_{s}}$ relations. We find that $M_\star = 0.507\pm 0.011\, M_\odot$ and $R_\star = 0.509\pm 0.015\, R_\odot$. Our derived stellar mass and radius are consistent with the values from \citetalias{Canas_2022} at $\leq 1\sigma$. We use our stellar mass and radius values to refine the planetary mass, semi-major axis, instellation, and equilibrium temperature of the hot Jupiter TOI-3714 b. The stellar and planetary parameters from \citetalias{Canas_2022} and from our work are summarized in Table~\ref{tab:star}.

\begin{deluxetable*}{lcccc}
\tablewidth{0pt}
\tablecaption{Stellar and planetary parameters of the TOI-3714 and TOI-5293 systems. \label{tab:star}}
\tablehead{
\colhead{Parameter} & \colhead{\citetalias{Canas_2022}} & \colhead{This work} & \colhead{\citetalias{Canas_2023}} & \colhead{This work} \\
&& (if different) && (if different)}
\startdata
& \multicolumn{2}{c}{\emph{TOI-3714 Stellar Parameters}} & \multicolumn{2}{c}{\emph{TOI-5293 A Stellar Parameters}} \\
& \multicolumn{2}{c}{(\emph{TIC 155867025, Gaia DR3 178924390478792320})} & \multicolumn{2}{c}{(\emph{TIC 250111245, Gaia DR3 2640121486388076032})} \\
$K_{s}$ &$10.852\pm 0.017$ & - & $11.639\pm 0.035$ & $11.825\pm 0.042$ \\
$M_{K_{s}}$ &$5.584\pm 0.018$& - & $5.589\pm 0.036$ & $5.775\pm 0.043$ \\
\teff{} [K] & $3660\pm 90$ & - & $3586\pm 88$ & - \\
$[\mathrm{Fe/H}]$ [dex] & $+0.1\pm 0.1$ & - & $-0.03\pm 0.12$ & - \\
\logg{} [dex] & $4.75\pm 0.05$ & $4.73\pm0.03$ & $4.77\pm0.05$ & $4.73\pm 0.03$ \\
$M_\star$ [$\mathrm{M}_{\odot}$] & $0.53\pm 0.02$ & $0.507\pm 0.011$ & $0.54\pm 0.02$ & $0.476\pm 0.012$ \\
$R_\star$ [$\mathrm{R}_{\odot}$] & $0.51\pm 0.01$ & $0.509\pm 0.015$ & $0.52^{+0.02}_{-0.01}$ & $0.479\pm 0.016$ \\ 
\prot{} [days] & $23.3\pm 0.3$ & - & $20.6^{+0.3}_{-0.4}$ & - \\
$v_{\mathrm{eq}}$ [km/s] & $1.10^{+0.03}_{-0.02}$ & $1.10\pm 0.04$ & $1.27\pm 0.04$ & $1.18\pm 0.04$ \\
\vsini{} [km/s] & $<2$ & $1.06^{+0.15}_{-0.12}$\tablenotemark{a} & $<2$ & $1.13^{+0.17}_{-0.10}$\tablenotemark{a} \\
\hline
& \multicolumn{2}{c}{\emph{TOI-3714 White Dwarf Primary Parameters}} & \multicolumn{2}{c}{\emph{TOI-5293 B Stellar Parameters}} \\
& \multicolumn{2}{c}{(\emph{TIC 662037581, Gaia DR3 178924390476838784})} &\multicolumn{2}{c}{(\emph{TIC 2052711961, Gaia DR3 2640121482094497024})} \\
$K_{s}$ &- & $17.40\pm 0.18$ & - & $13.65\pm 0.13$ \\
$M_{K_{s}}$ &- & $12.14\pm 0.19$& - & $7.60\pm 0.13$ \\
\teff{} [K] &-&-& $3041^{+280}_{-41}$ & - \\
\logg{} [dex] &-&-& $4.72^{+0.16}_{-0.14}$ & $5.00\pm 0.06$ \\
$M_\star$ [$\mathrm{M}_{\odot}$] & $\approx 1.07$ & - & - & $0.214\pm 0.015$ \\
$R_\star$ [$\mathrm{R}_{\odot}$] &-&-& $0.26^{+0.10}_{-0.08}$ & $0.244\pm 0.015$ \\ 
\hline
& \multicolumn{2}{c}{\emph{TOI-3714 b Planetary Parameters}} & \multicolumn{2}{c}{\emph{TOI-5293 A b Planetary Parameters}} \\
$P$ [days] &$2.154849\pm 0.000001$& - & $2.930289\pm 0.000004$ & -  \\
$T_0$ [BJD] &$2458840.5093\pm 0.0004$ & - & $2459448.9148\pm 0.0004$ & -  \\
$T_{14}$ [hours] & $1.66^{+0.02}_{-0.01}$ & - & $1.94^{+0.05}_{-0.04}$ & - \\
$R_p/R_\star$  &$0.204\pm 0.003$& - & $0.210^{+0.005}_{-0.004}$ &-  \\
$a/R_\star$ &$11.5^{+0.4}_{-0.5}$ & - & $14.1^{+1.6}_{-1.1}$ & - \\
$b$ &$0.26^{+0.08}_{-0.10}$ & - & $0.32^{+0.12}_{-0.14}$ & - \\
$K$ [m/s] &$169^{+6}_{-5}$ & - & $115.6\pm 14.5$ & -  \\
$\sqrt{e}\cos{\omega}$ &$0.0\pm 0.1$ & - & $-0.07^{+0.17}_{-0.16}$ & -  \\
$\sqrt{e}\sin{\omega}$ & $0.1\pm 0.1$ & - & $-0.17\pm 0.22$ & - \\
$e$ &$0.03^{+0.03}_{-0.02}$ & - & $<0.38$\tablenotemark{b} & - \\
$R_p$ [$\mathrm{R}_{J}$] &$1.01\pm 0.03$& - & $1.06\pm 0.04$ & $0.98\pm 0.04$  \\
$M_p$ [$\mathrm{M}_{J}$] & $0.70\pm 0.03$& $0.67\pm 0.02$ & $0.54\pm 0.07$ & $0.50\pm 0.07$  \\
$a$ [au] &$0.027\pm 0.001$ & $0.0260\pm 0.0002$ & $0.034^{+0.004}_{-0.003}$ & $0.0313\pm 0.0003$ \\    
$S$ [$S_\oplus$] &$54^{+4}_{-5}$& $62\pm 7$ & $34.6^{+8.6}_{-6.2}$ & $34.8^{+4.4}_{-4.0}$  \\    
$T_{\mathrm{eq}}$ [K]\tablenotemark{c} &$750\pm 20$& $781\pm 23$ & $675^{+42}_{-30}$ & $676\pm 20$  \\    
\enddata
\tablenotetext{a}{We measure \vsini{} from the Rossiter-McLaughlin effect. These results are presented in Section~\ref{subsect:results}.}
\tablenotetext{b}{99\% upper limit.}
\tablenotetext{c}{Equilibrium temperature assuming zero albedo and perfect heat redistribution.}
\end{deluxetable*}

\subsection{The S-Type Planetary System around TOI-5293 A} \label{sect:system5293}
\citealt{Canas_2023} (hereafter \citetalias{Canas_2023}) reported the discovery of a hot Jupiter on an S-type orbit around the primary star in the TOI-5293 binary system. The primary star, TOI-5293 A, is an M3 dwarf while the companion star is a widely separated M4 dwarf with an angular separation of 3\farcs57 and a projected separation of 574 au. The binary system is located at a distance of $162.2\pm 0.7$ pc.

Following the methodology of \citetalias{Canas_2022} for TOI-3714, \citetalias{Canas_2023} report a model-dependent stellar mass and radius for TOI-5293 A based on fits of the stellar SED to MIST isochrones. Following our procedure to rederive the stellar mass and radius of TOI-3714 in Sect.~\ref{sect:system3714}, we refine the stellar mass and radius of TOI-5293 A after deriving the dilution correction for the $K_s$-band magnitude of TOI-5293 A. The fainter M4 companion TOI-3714 B does not appear in the 2MASS point source catalog and the 2MASS photometric quality flags for the primary star are all nominal, implying that reported $K_s$-band photometry for TOI-5293 A is actually the combined flux of both stellar components. Using the stars' Gaia photometry ($\Delta G = 2.90$), we estimate the $K_s$-band flux ratio $F_{K_s,A}/F_{K_s,B}$ to be $5.40 \pm 0.74$, which corresponds to a dilution factor of $F_{K_s,A}/(F_{K_s,A}+F_{K_s,B})=0.842\pm 0.018$. Taking this effect into account, we revise the $K_s$-band magnitude of TOI-5293 A from $K_s = 11.639\pm  0.035 \to 11.825\pm  0.042$. Plugging the revised $K_s$-band magnitude and parallax from Gaia DR3 into the aforementioned mass-$M_{K_s}$ and radius-$M_{K_s}$ relations, we derive the stellar mass and radius of TOI-5293 A of $0.476\pm 0.012\, M_\odot$ and $0.479\pm 0.016\, R_\odot$, respectively. 

We highlight that our revised values are significantly smaller than the values of $0.54\pm 0.02\, M_\odot$ and $0.52^{+0.02}_{-0.01}\, R_\odot$ reported by \citetalias{Canas_2023}. Those values do not include photometric dilution corrections and are derived from model-dependent isochrone fits to the stellar SED. We use our stellar mass and radius values to refine the planetary radius, mass, semi-major axis, instellation, and equilibrium temperature of the hot Jupiter TOI-5293 A b. The stellar and planetary parameters from \citetalias{Canas_2023} and our work are compared in Table~\ref{tab:star}.

Using our refined $K_s$-band magnitude for TOI-5293 A and the $K_s$ magnitude difference between the two stars, we estimate TOI-5293 B to have $K_s=13.65\pm 0.13$. Recall that TOI-5293 B does not have any existing 2MASS photometric measurements due to blending with TOI-5293 A. The corresponding stellar mass and radius are $0.214\pm 0.015\, M_\odot$ and $0.244\pm 0.015\, R_\odot$, respectively.


\section{Transit Observations with MAROON-X} \label{sect:data}
We observed one full transit of TOI-3714 b on December 10, 2024 UT 
using the MAROON-X spectrograph under the program GN-2024B-Q-132 (PI: Cloutier). MAROON-X is a high-resolution ($R\sim 85,000)$ optical echelle spectrograph located at the 8m Gemini-North Telescope on Maunakea, Hawai'i \citep{seifahrt18,Seifahrt2020}. Our observations were taken in the spectrograph's blue ($500-670$ nm) and red ($650-900$ nm) arms simultaneously. The continuous observing sequence spanned 4.18 hours and consisted of sixteen 900-second exposures. The observing sequence covered airmasses between 1.06-1.24 and we achieved median peak signal-to-noise ratios (S/N) of 14.5 and 27 in the blue and red arms, respectively. 

We also observed two full transits of TOI-5293 A b on July 3, 2024 UT and September 29, 2024 UT under the programs GN-2024A-Q-131 (PI: Cloutier) and GN-2024B-Q-132 (PI: Cloutier). Our observing setups were identical to those described above for TOI-3714. The continuous transit observing sequences contained thirteen and seventeen measurements spanning 3.33 hours and 4.45 hours, respectively. During the first transit observation, the target rose from an airmass of 2.37 to 1.10, and we achieved median peak S/N values of 17 and 31, respectively. The second transit sequence covered airmasses between 1.08-1.49 and we achieved median peak S/N values of 19 and 33 in the blue and red arms, respectively. 

All our raw MAROON-X data were reduced using custom routines originally written for VLT/CRIRES \citep{Bean_2010}. We extracted RV measurements from both the blue and red arms using the \texttt{SERVAL} template-matching code \citep[\texttt{Spectrum Radial Velocity Analyzer};][]{Zechmeister_2018}. \texttt{SERVAL} also extracts the following stellar activity metrics: the chromatic RV index (CRX), differential line widths (dLW), and $H\alpha$ activity indices \citep{Zechmeister_2018}. Tables of our time series are provided in Appendix~\ref{app:tables}. Our RV extractions yielded median blue, red, and combined RV uncertainties for TOI-3714 of 8.57 m/s, 5.61 m/s, and 4.69 m/s, respectively. Similarly for Transit 1 of TOI-5293 A b: 6.62 m/s, 4.40 m/s, and 3.65 m/s. Similarly for Transit 2 of TOI-5293 A b: 5.44 m/s, 4.24 m/s, and 3.34 m/s, respectively.

\section{Data Analysis} \label{sect:analysis}
\subsection{Identifying detrending variables} \label{sect:variables}
\begin{figure}
    \centering
    \includegraphics[width=\hsize]{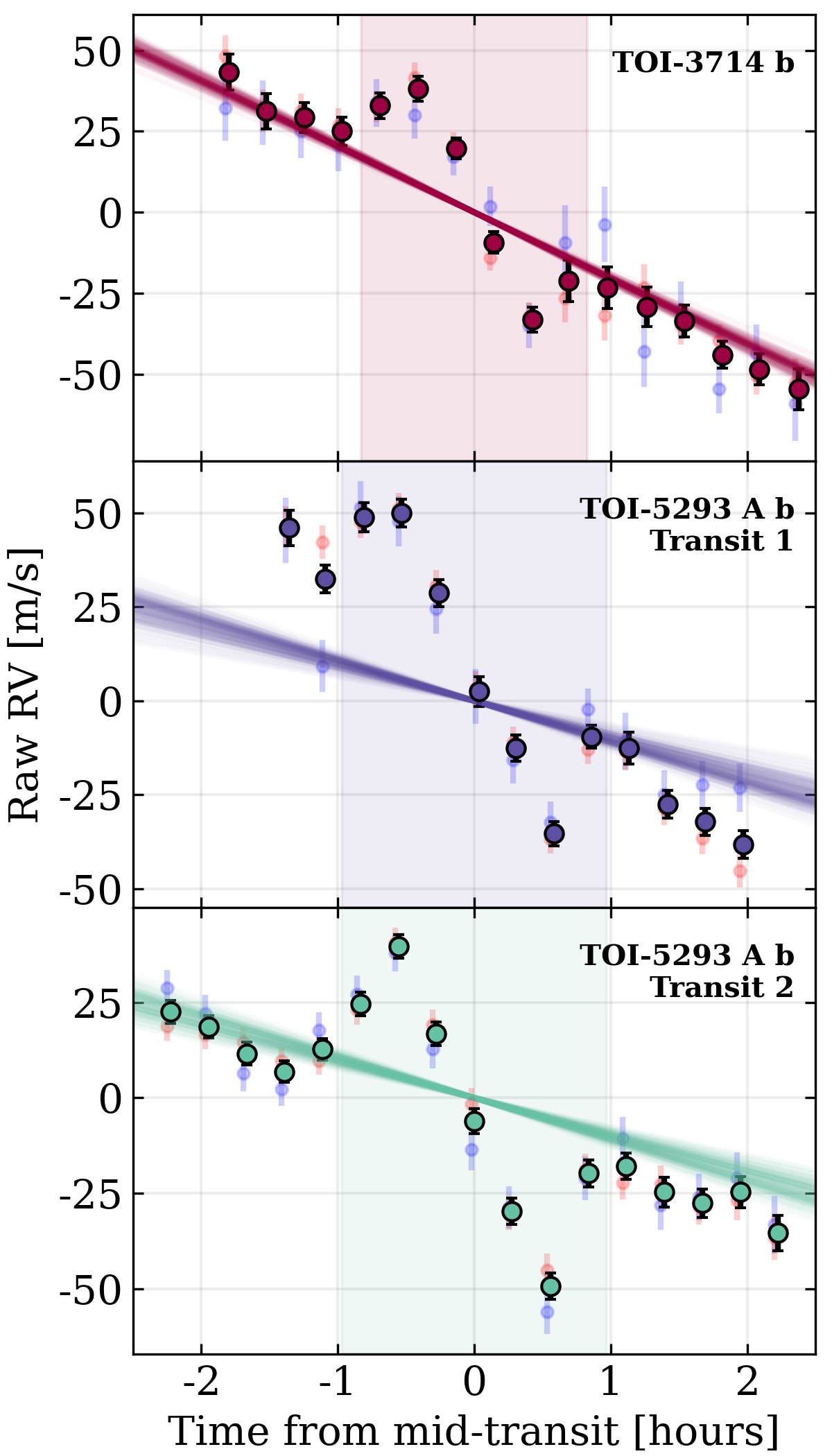}
    \caption{Our raw MAROON-X data of TOI-3714 b (top panel) and TOI-5293 A b (two lower panels). The blue arm, red arm, and combined RVs are depicted by the translucent blue, red, and solid markers, respectively. The solid lines depict random draws from each planet's known Keplerian orbit posterior (TOI-3714 b; \citetalias{Canas_2022}, TOI-5293 A b; \citetalias{Canas_2023}). The vertical shaded regions highlight each planet's in-transit window.}
    \label{fig:rawrv}
\end{figure}

We visually inspected the raw RV data for each transit observing sequence (Figure~\ref{fig:rawrv}). The Keplerian signals induced by each planet are known from their respective discovery papers. After removing the Keplerian signals we identified a significant and nearly linear trend in the blue and red arm time series of Transit 1 of TOI-5293 A b. The amplitude of this linear trend was in excess of the known Keplerian motion, which prompted the investigation of correlations between the blue and red arm RVs with other contemporaneous observables. The two remaining transit sequences did not show similarly significant trends. 

We calculated the pearson correlation coefficients of the out-of-transit blue and red arm RV data to search for correlations with the following quantities: airmass, barycentric Earth radial velocity (BERV), the chromatic index (CRX), and differential line width (dLW). For all transits, the correlations with the blue arm RVs are systematically weaker than in the red arm, likely due to their elevated rms relative to the red arm RVs. We found that for the first transit of TOI-5293 A b, there exists a strong correlation between the Keplerian-subtracted RVs and the BERV ($\rho=0.995$). Noting that the BERV evolves approximately linearly over the 3.33-hour observing baseline, we elect to include a linear detrending term as a function of time in our full RV model. For consistency, we include similar detrending terms in our models of all observed transits by allowing the linear detrending slope values to go zero if favored by the data. The exact cause of this drift remains unknown, although a recent investigation of data-driven RV extraction methods, including template-matching, has shown that these methods are capable of producing quasi-linear accelerations in time series lasting a few hours \citep{Silva_2025}. Despite the unknown origin of this trend, in the following section we will show that a linear detrending term is sufficient to model the data.

\subsection{Full RV model} \label{sect:rvmodel}
We use the \texttt{starry} package \citep{Luger_2019} to compute forward models of the classical RM effect and Keplerian RV signals. The classical RM effect is agnostic to spatial variations in the host star's surface RV field that do not arise from rigid-body rotation \citep{Cegla_2016,Bourrier_2024}. Within \texttt{starry}, the classical RM effect is modeled by a star whose surface brightness is initially uniform, as parameterized by the spherical harmonic with degree and order $\{l,m\}=\{0,0\}$, but also includes quadratic limb-darkening. Due to the overlapping wavelength coverage of TESS and MAROON-X, we adopt the TESS-band limb-darkening parameters from each planet's respective discovery paper. The surface RV field is generated by rotating the star with velocity $v_{\mathrm{eq}}$, with an inclination relative to the observer's line-of-sight $i_\star$, and a sky-projected stellar obliquity $\lambda$. The in-transit RM signature is calculated by evolving a planet with size $R_p/R_\star$ on a Keplerian orbit. This parameterization within \texttt{starry} is analogous to analytical prescriptions of the classical RM effect from the literature \citep{Ohta_2005, Hirano_2010, Hirano_2011}.

Our full RV model features the following twelve parameters: an RV offset $v_0$, a five-parameter Keplerian orbit parameterized by the planet's orbital period $P$, time of mid-transit $T_0$, RV semi-amplitude $K$, $h=\sqrt{e}\cos{\omega}$, and $k=\sqrt{e}\sin{\omega}$, where $e$ and $\omega$ are the Keplerian orbital eccentricity and argument of periastron, respectively, the classical RM effect parameterized by the planet-to-star radius ratio $R_p/R_\star$, the transit impact parameter $b$, the projected stellar rotation velocity \vsini{,} and the projected stellar obliquity $\lambda$, plus a scalar jitter term $\log{\sigma}$. As discussed in Section~\ref{sect:variables}, we include the slope of a linear detrending component $m$, which is most significant for our first transit observation of TOI-5293 A b (c.f. Figure~\ref{fig:rawrv}). 

We also note that our model ignores the effects of stellar activity. M dwarf activity at field ages is dominated by rotationally-modulated magnetically active regions (AR). These stars can also exhibit flares, albeit with a much lower frequency than M dwarfs younger than $\sim 800$ Myr \citep{Feinstein_2020}. We inspected our H$\alpha$ time series and confirmed that no transit sequence exhibited any stochastic variability consistent with a flare. We also note that our observations are largely insensitive to unocculted ARs due to the baseline of each transit observation being small compared to the stars' rotation periods of 20.6-23.3 days (c.f. Table~\ref{tab:star}). While occulted ARs could impact our observations, we cannot reliably assess the occurrence of an occulted AR with our MAROON-X data, which would benefit from simultaneous photometry. We note that we expect the impact of occulted starspots and plages to be negligible on both of our low activity stars as neither star exhibits significant levels of photometric or RV variability. 

\begin{figure*}
\centering
\includegraphics[width=0.495\hsize]{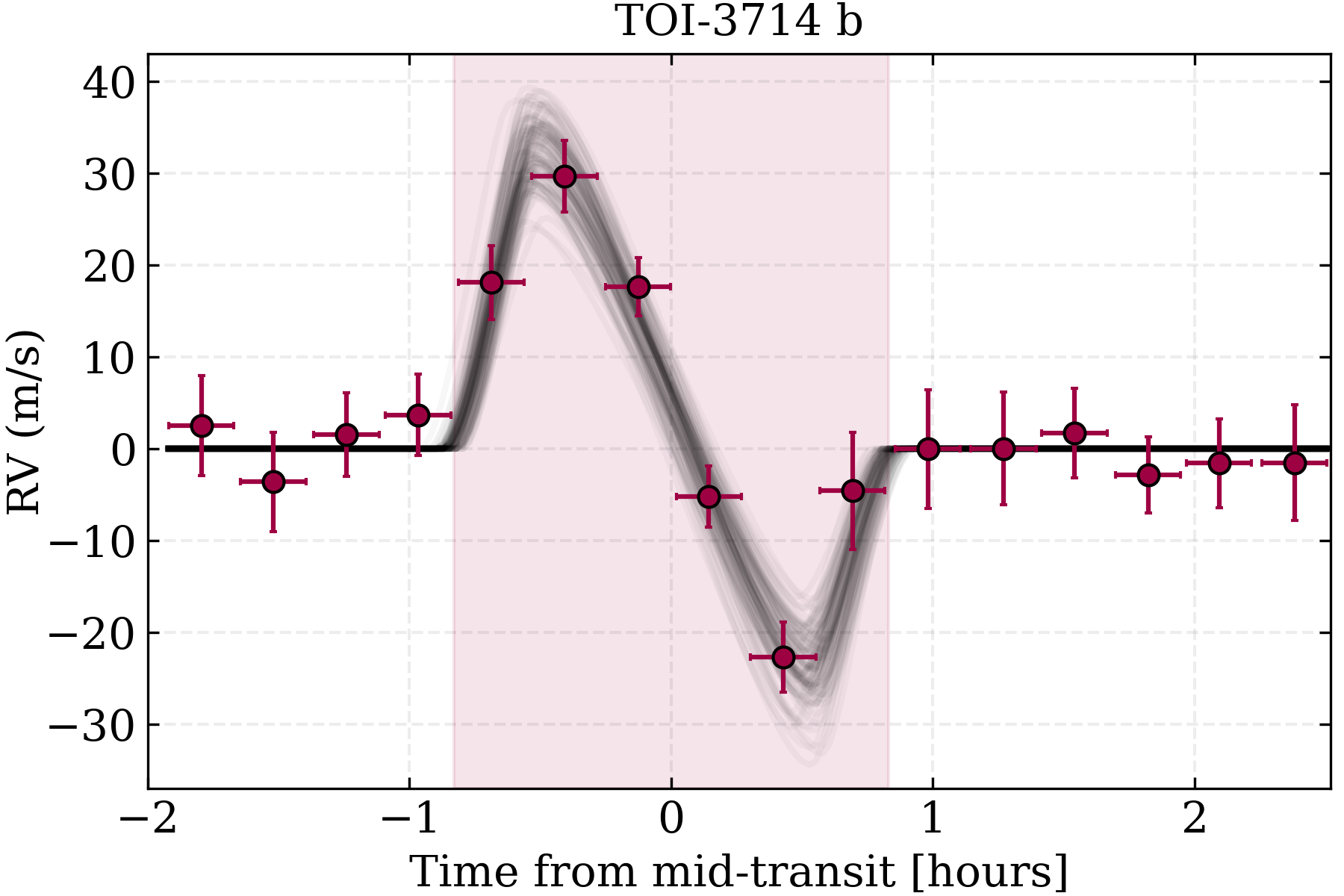}
\includegraphics[width=0.495\hsize]{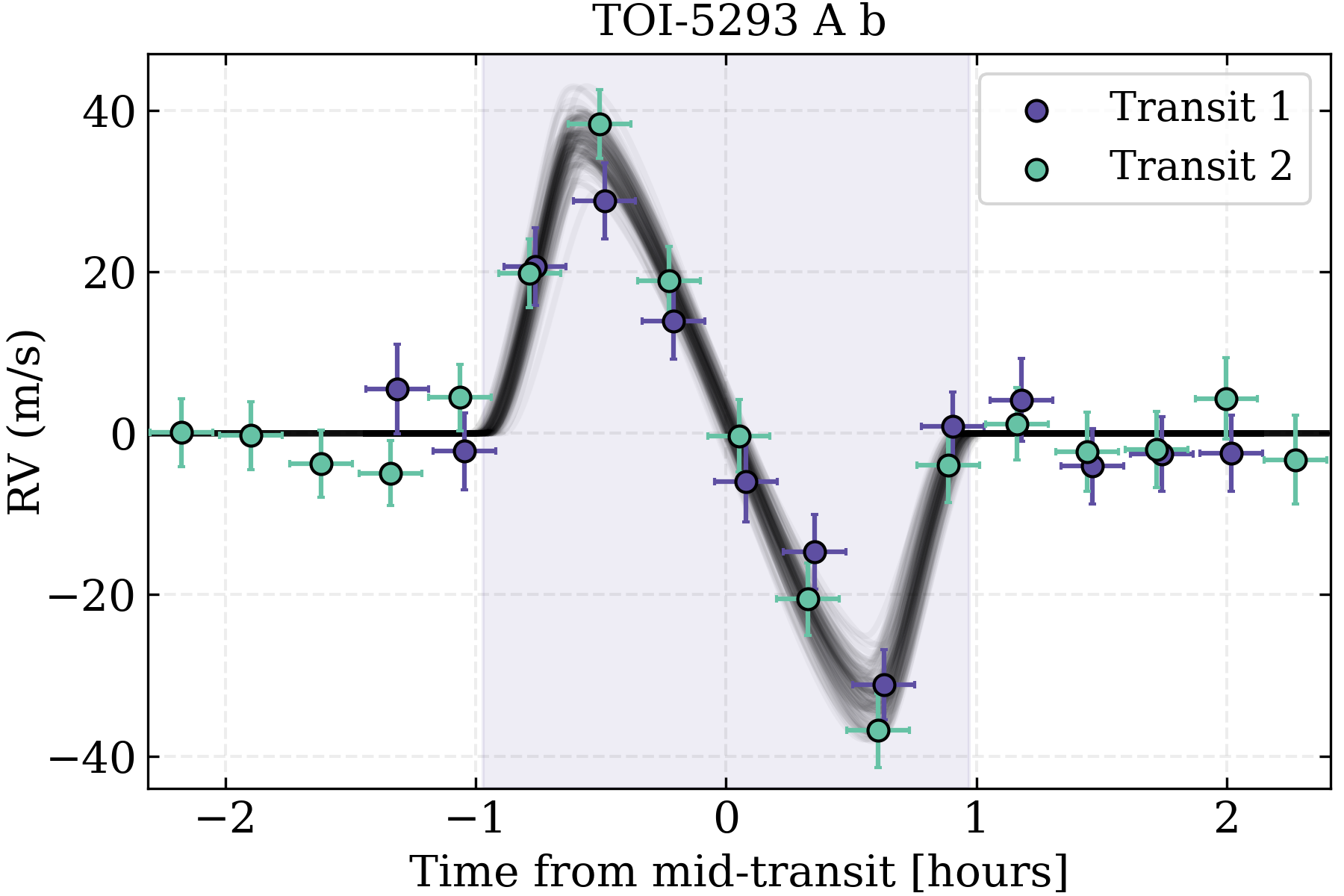}
\caption{The RM effect signatures for TOI-3714 b (left panel) and TOI-5293 A b (right panel). Each set of markers depicts the combined MAROON-X RV observations (i.e. blue plus red channels) for each transit sequence observed. The black curves represent random draws from our RM model posteriors after removing the median detrending model for each transit and the planet-induced Keplerian orbital signals.}
\label{fig:rm}
\end{figure*}

The short durations of our transit observations with MAROON-X provide very weak independent constraints on the Keplerian and transit parameters compared to the inferences from each planet's transit and orbital RV time series. We therefore need not reproduce those analyses by \citetalias{Canas_2022} and \citetalias{Canas_2023} for TOI-3714 and TOI-5293 A, respectively, and instead adopt the parameter posteriors for $\{ P, T_0, K, h, k, R_p/R_\star, b \}$ from those studies as priors. The MAROON-X data presented herein therefore provide novel constraints on the parameters $\{ v_0, v\sin{i_\star}, \lambda, \log{\sigma}, m \}$. Our model parameter priors are reported in Table~\ref{tab:results}.

We use the \texttt{emcee} Markov chain Monte Carlo (MCMC) package \citep{foremanmackey13} to sample the posteriors of our full RV models. We use 64 walkers to sample 60,000 steps for each target. Our chain length corresponds to $\sim 30\times$ the autocorrelation time and we omit the first 20\% of steps as burn-in. When evaluating the model at each step in the MCMC chain, we oversample the model by a factor of 20 over a range of times centered on each epoch of observation $t_{\mathrm{obs},i}$. We then compute the average of the model to calculate the likelihood function. Specifically, we evaluate the model at $t_{\mathrm{obs},i}\pm t_{\mathrm{exp}}/2$, where $t_{\mathrm{exp}}=900$ seconds. This process is necessary to capture the sharp inflections in the RM model that occur on time scales $\lesssim t_{\mathrm{exp}}$ (i.e. shortly after transit ingress and shortly before transit egress; c.f. Figure~\ref{fig:rm}).

\section{Discussion} \label{sect:discussion}
\subsection{Results} \label{subsect:results}
Figure~\ref{fig:rm} depicts the results of our RV modeling for TOI-3714 b and TOI-5293 A b. Point estimates of our fitted model parameters are reported in Table~\ref{tab:results}.

We detect the RM effect for both planets at high significance with semi-amplitudes of $28.8\pm 3.8$ m/s ($7.6\sigma)$ and $31.4\pm 4.1$ m/s ($7.8\sigma)$ for TOI-3714 b and TOI-5293 A b, respectively. We measure projected stellar obliquities of $\lambda = 21^{+14}_{-11} \si{\degree}$ and $-12^{+19}_{-14} \si{\degree}$, and projected stellar rotation velocities of $v\sin{i_\star} = 1.06^{+0.15}_{-0.12}$ km/s and $1.13^{+0.17}_{-0.10}$ km/s for TOI-3714 and TOI-5293 A b, respectively. Our measurements represent just the second and third detections of the RM effect for HJMDs.

\begin{deluxetable*}{lcccc}
\tablewidth{0pt}
\tablecaption{Model parameter priors and posterior point estimates. \label{tab:results}}
\tablehead{\colhead{Parameters} & \multicolumn{2}{c}{TOI-3714} & \multicolumn{2}{c}{TOI-5293 A} \\
& Prior & Posterior & Prior & Posterior}
\startdata
& \multicolumn{4}{c}{\emph{Observed parameters}} \\
\vspace{-0.2cm} &&&& $-9^{+10}_{-12}$\tablenotemark{a} \\
\vspace{-0.2cm} Velocity offset, $v_0$ [m/s] & $\mathcal{U}(-50,50)$ & $3^{+9}_{-12}$ & $\mathcal{U}(-50,50)$ &  \\
&&&& $-24^{+10}_{-13}$ \\
\vspace{-0.2cm} &&&& $-12.6\pm2.0$\tablenotemark{a} \\
\vspace{-0.2cm} Linear slope, $m$ [m/s/hr] & $\mathcal{U}(-50,50)$ & $-2.7^{+2.2}_{-2.4}$ & $\mathcal{U}(-50,50)$ &  \\
&&&& $-1.4^{+1.8}_{-1.7}$ \\
Orbital period, $P$ [days] & $\mathcal{N}(2.154849,0.000001)$ & $2.154850\pm 0.000001$ & $\mathcal{N}(2.930289,0.000004)$ & $2.930287\pm 0.000004$ \\
\vspace{-0.2cm} Time of mid-transit, &&&& \\ 
\vspace{-0.3cm}
& $\mathcal{N}(1840.5093,0.0004)$ & $1840.5093\pm 0.0004$ & $\mathcal{N}(2448.9148, 0.0004)$ & $2448.9147\pm 0.0004$ \\
$T_0$ [BJD-2,457,000] &&&& \\
RV semi-amplitude, $K$ [m/s] & $\mathcal{SN}(169,6,5)$ & $167\pm 5$ & $\mathcal{N}(115.6,14.5)$ & $115\pm 14$ \\
$h=\sqrt{e}\cos{\omega}$ & $\mathcal{N}(0.0,0.1)$ & $0.00\pm 0.08$ & $\mathcal{SN}(-0.07,0.17,0.16)$ & $-0.04\pm 0.16$ \\
$k=\sqrt{e}\sin{\omega}$ & $\mathcal{N}(0.1,0.1)$ & $0.01\pm 0.12$ & $\mathcal{N}(-0.17,0.22)$ & $-0.03\pm 0.21$ \\
\vspace{-0.2cm} Planet-to-star radius ratio, &&&& \\
\vspace{-0.3cm} 
& $\mathcal{N}(0.204,0.003)$ & $0.204\pm 0.003$ & $\mathcal{SN}(0.210,0.005,0.004)$ & $0.210\pm 0.005$ \\
$R_p/R_\star$ &&&& \\
Impact parameter, $b$ & $\mathcal{SN}(0.26,0.08,0.10)$ & $0.27^{+0.09}_{-0.10}$ & $\mathcal{SN}(0.32,0.12,0.14)$ & $0.15^{+0.11}_{-0.10}$ \\
\vspace{-0.2cm} Projected stellar rotation &&&& \\
\vspace{-0.3cm}
& $\mathcal{U}(0,5)$ & $1.06^{+0.15}_{-0.12}$ & $\mathcal{U}(0,5)$ & $1.13^{+0.17}_{-0.10}$ \\
velocity, \vsini{} [km/s] &&&& \\
\vspace{-0.2cm} Projected stellar obliquity, &&&&\\
\vspace{-0.3cm} 
& $\mathcal{U}(-180,180)$ & $21^{+14}_{-11}$ & $\mathcal{U}(-180,180)$ & $-12^{+19}_{-14}$ \\
$\lambda$ [deg] &&&& \\
Jitter factor, $\log{\sigma} [m/s]$ & $\mathcal{U}(-1,1)$ & $0.15^{+0.05}_{-0.07}$ & $\mathcal{U}(-1,1)$ & $0.47^{+0.02}_{-0.05}$ \\
& \multicolumn{4}{c}{\emph{Derived parameters}} \\
Stellar inclination, $i_\star$ [deg] & $-$ & $76.1^{+8.9}_{-8.1}$ & $-$ & $80.5^{+6.5}_{-7.8}$ \\
\vspace{-0.2cm} Deprojected stellar obliquity, &&&& \\
\vspace{-0.3cm}
& $-$ & $26^{+11}_{-10}\: (<42)$\tablenotemark{b} & $-$ & $24^{+11}_{-10} (<40)$\tablenotemark{b} \\
$\psi$ [deg] &&&& \\
\enddata
\tablenotetext{a}{The upper and lower values correspond to the observed transits of TOI-5293 A b \#1 and \#2, respectively.}
\tablenotetext{b}{The values reported in parentheses represent the $95^{\mathrm{th}}$ percentiles.}
\end{deluxetable*}

\begin{figure*}
\centering
\includegraphics[width=\hsize]{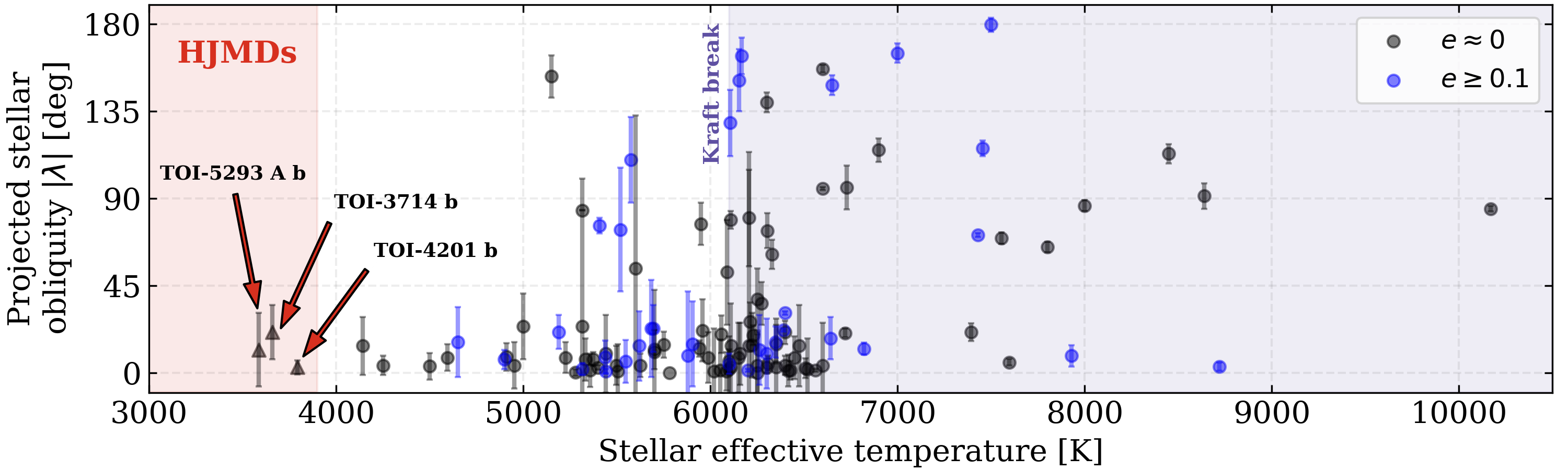}
\caption{The distribution of projected stellar obliquities $|\lambda|$ for giant exoplanets ($R_p\geq 5\, R_\oplus$) as a function of host star effective temperature. All stellar obliquity measurements come from measurements of the Rossiter-McLaughlin effect and are retrieved from the NASA Exoplanet Archive. Hot Jupiters around M dwarfs (HJMDs) occupy the shaded red region of parameter space with \teff{} $<3900$ K, while host stars above the Kraft break (\teff{} $\geq 6100$ K) occupy the shaded blue region. The HJMDs TOI-3714 b and TOI-5293 A b from this work, and TOI-4201 \citep{Gan_2024} are highlighted by red arrows.}
\label{fig:spinorbit}
\end{figure*}

Figure~\ref{fig:spinorbit} compares our projected stellar obliquity measurements to the distribution of stellar obliquities of gas giant hosts as a function of host star effective temperature. These data were retrieved from the NASA Exoplanet Archive \citep{Akeson_2013}. The projected stellar obliquities of all three HJMD hosts (i.e. TOI-3714, TOI-5293 A, and TOI-4201; \citealt{Gan_2024}) are consistent with being aligned. Given that these planets' planet-to-star mass ratios span $\sim 1 \times 10^{-3} -4 \times 10^{-3}$, their aligned obliquities are consistent with recent empirical results that suggest that massive planets with $M_p/M_\star \gtrsim 3\times 10^{-4}$ around stars below the Kraft break \citep[i.e. \teff{} $< 6100$ K;][]{Kraft_1967} are preferentially tidally realigned \citep{Rusznak_2025}. While a sample of three HJMD obliquities lies firmly within the regime of small number statistics, in Section~\ref{sect:formation} we will present a preliminary discussion of the implications that all three HJMDs with RM detections have $|\lambda|$ consistent with zero. Future RM measurements of more HJMDs will continue to improve comparisons of the HJMD obliquity distribution to HJs around earlier type stars.



\subsection{Deprojected stellar obliquities} \label{sect:psi}
While the RM effect is sensitive to the sky-projected stellar obliquity $\lambda$, inferences on the dynamical histories of individual transiting planets require the deprojected (i.e. 3D) stellar obliquity $\psi$. The 3D stellar obliquity is calculated from 

\begin{equation}
\cos{\psi} = \cos{i_\star} \cos{i_p} + \sin{i_\star} \sin{i_p} \cos{\lambda},
\label{eq:psi}
\end{equation}

\noindent where $i_\star$ is the inclination angle of the host star's rotation axis, $i_p$ is the orbital inclination of the transiting planet, and $\lambda$ is the projected stellar obliquity from the RM effect \citep{Fabrycky_2009}. We recover the values of $i_\star$ for our targets using a-priori knowledge of their stellar radii and photometric rotation periods (c.f. Table~\ref{tab:star}) along with our measurements of $v\sin{i_\star}$ from the RM effect (Table~\ref{tab:results}). We follow the MCMC-based methodology of \cite{Masuda_2020} to probabilistically infer $\cos{i_\star}$ from independent datasets that measure $v_\mathrm{eq} = 2\pi R_\star/P_{\mathrm{rot}}$ and $v\sin{i_\star}$, respectively. We derive $\cos{i_\star}=0.24^{+0.14}_{-0.15}$ ($i_\star=76.1^{+8.9}_{-8.1} \si{\degree}$) and 
$\cos{i_\star}=0.17^{+0.13}_{-0.11}$ 
($i_\star=80.5^{+6.5}_{-7.8} \si{\degree}$) for TOI-3714 and TOI-5293 A, respectively. Plugging these values into Eq.~\ref{eq:psi} yields deprojected stellar obliquities of $\psi=26^{+11}_{-10} \si{\degree}$ and $24^{+11}_{-10} \si{\degree}$ for TOI-3714 and TOI-5293 A, respectively. While both systems show evidence for a mild 3D misalignment at $\sim 2\sigma$, it is clear that neither planet exhibits a significant misalignment such that we consider them to be well-aligned. We note that this result could be refined with additional transit observations to improve the sampling of the RM feature and yield tighter constraints on its amplitude, and consequently on \vsini{} and $\psi$.

\subsection{Kozai-Lidov in the TOI-3714 and TOI-5293 planetary systems}
Planets with S-type orbits in a stellar binary will be acted on by the Kozai-Lidov mechanism \citep[KL;][]{Kozai_1962,Lidov_1962} if the initial inclination between the orbital planes of the planet and the binary companion is sufficiently large\footnote{We note that although the KL mechanism will always act on planets with S-type orbits, below a minimum mutual inclination the planet's argument of periapse will circle through all angles rapidly and effectively average out the forced eccentricity by the binary companion to zero.} \citep[i.e. $\gtrsim 39.2 \si{\degree}$ for a planetary test particle;][]{Innanen_1997}. The KL mechanism is a secular angular momentum exchange process between the planet and binary companion that produces large periodic oscillations between the planet's eccentricity $e_p$ and relative inclination $\Delta i$ that conserves the Kozai integral $\sqrt{1-e_p^2}\cos{\Delta i}$. Given an initial mutual inclination $\Delta i_0$, the planet's maximum eccentricity that can be achieved is $e_{p,\mathrm{max}} \approx \sqrt{1 - 5/3\cos^2{\Delta i_0}}$ \citep{Innanen_1997}. The combined effects of the KL mechanism and tidal circularization at small pericenter distances (i.e. when $e_p\approx e_{p,\mathrm{max}}$) is known as KL migration, which may be partly responsible for the observed close-in orbits of HJs \citep{Wu_2003}.

Here we investigate whether the binary companions in the TOI-5293 and TOI-3714 planetary systems could plausibly drive KL migration of the systems' HJs. We posit that KL migration is likely to operate if the following conditions are met: 1) the minimum mutual inclination of the planet and binary companion is large enough for the KL mechanism to operate and 2) the timescale of KL oscillations is short compared to the age of the system.

\begin{figure*}
\centering
\includegraphics[width=0.9\hsize]{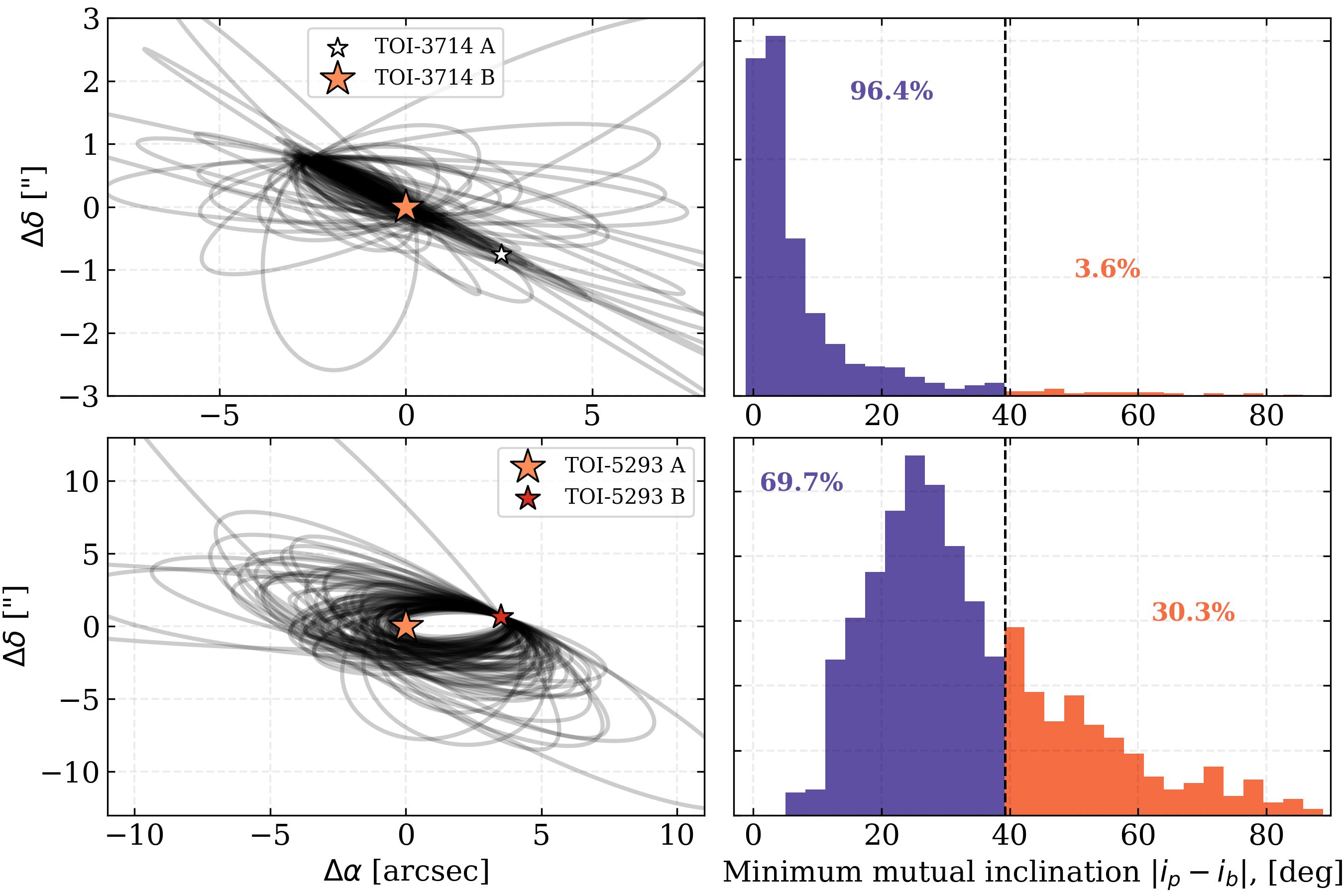}
\caption{Results of orbital fits to Gaia DR3 astrometric data for the binary systems TOI-3714 (upper row) and TOI-5293 (lower row). Left column: a random selection of sky-projected binary orbit solutions from our astrometric fit posteriors. Right column: the distribution of minimum mutual inclinations between each system's planet and binary companion. The vertical dashed line at $\arccos{\sqrt{3/5}}\approx 39.2 \si{\degree}$ corresponds to the minimum mutual inclination needed to drive KL migration. The percentiles of the mutual inclination distribution above and below $39.2 \si{\degree}$ are depicted by different colors and are annotated for each system.}
\label{fig:lofti}
\end{figure*}

We begin by constraining the orbital inclinations of the binary star systems using the system's Gaia DR3 astrometry \citep{GaiaDR3}. We use the \texttt{lofti\_gaia} package \citep{Blunt_2017,Pearce_2020} to fit the Keplerian orbital parameters of each binary system given input astrometry and the masses for both stellar components (see Table~\ref{tab:star}). Figure~\ref{fig:lofti} depicts a selection of orbits in the sky plane based on random samples from the joint posterior of Keplerian orbital parameters. Figure~\ref{fig:lofti} also depicts each system's distribution of minimum mutual inclinations between the orbital planes of the planet and binary companion $\Delta i = |i_p - i_b|$. We find that the TOI-5293 system exhibits a significant probability of being misaligned with the system's hot Jupiter, with $\sim 30$\% of the minimum mutual inclination posterior exceeding the minimum $\Delta i$ necessary to trigger the KL mechanism. Conversely, the astrometric parameters of the TOI-3714 system favor an orbital solution that may be aligned with the system's HJ with only $\sim 3$\% of the $\Delta i$ posterior exceeding $39.2 \si{\degree}$. We highlight that $\Delta i$ is the \emph{minimum} mutual inclination between the planetary and binary orbital planes because the orientation angles of the planetary orbits are unknown. Therefore, while the minimum mutual inclination of the TOI-3714 system is consistent with zero, this does not rule out that the system's true mutual inclination may be sufficiently misaligned to drive KL migration. 

We conclude that there is a high likelihood that the mutual inclination of the TOI-5293 system is sufficiently misaligned to drive KL migration. Conversely, we conclude that the mutual inclination of the TOI-3714 system may be sufficient to drive KL migration and that no data are available to make a more definitive statement. We also note the possibility that because TOI-3714 A is a white dwarf, the formation channel of the HJMD TOI-3714 b may be distinct because post-main sequence evolution in binary systems can alter the dynamics of S-type planets \citep[e.g.][and references therein]{Veras_2016} including the possibility of the planets ``hopping'' between stars \citep{Kratter_2012}.

The second condition for KL migration is that the KL oscillation timescale is less than the system's age. We calculate each system's KL oscillation timescale 

\begin{equation}
\tau_{\mathrm{KL}} \approx P_b \left( \frac{M_1 + M_2}{M_2} \right) \left( \frac{P_b}{P_p} \right) (1-e_b^2)^{3/2},
\end{equation}

\noindent where $M_1$ and $M_2$ are the masses of the binary's primary and secondary components, respectively, $P_b$ and $e_b$ are the binary's orbital period and eccentricity, respectively, and $P_p$ is the orbital period of the planet around the primary \citep{Kiseleva_1998}. We compute $\tau_{\mathrm{KL}}$ for each system using our $P_p$ posteriors (Table~\ref{tab:results}), the binary component masses listed in Table~\ref{tab:star}, and the $P_b$ posterior from our astrometric fitting. We set $e_b=0$ because our astrometric fitting provides only weak constraints on $e_b$ and because the maximum value of $\tau_{\mathrm{KL}}$ is achieved when $e_b=0$. 

We calculate KL oscillation timescales of $\tau_{\mathrm{KL}} = 5.7^{+15.1}_{-2.8}\times 10^3$ years and $6.0^{+9.4}_{-2.6}\times 10^4$ years for TOI-3714 and TOI-5293, respectively. While precise ages for these systems remain elusive, \citetalias{Canas_2022} constrain the age of TOI-3714 to be $0.7-5.1$ Gyrs based on the primary star's rotation period of 23.3 days and the rotation-kinematic age relation for M dwarfs from \cite{Newton_2016}. Following the same methodology, \citetalias{Canas_2023} infer TOI-5293 to be a field age system given the 20.6-day rotation period of the primary. Each system's approximate field age is approximately four orders of magnitude longer than their respective KL oscillation timescales. We conclude that both systems would have had sufficient time for KL oscillations to operate.


\subsection{Dynamically cold migration of HJMDs versus tidal damping} \label{sect:formation}
Aligned stellar obliquities of HJMD hosts are predicted by two scenarios: 1) primordial alignments resulting from dynamically cold migration (i.e. disk-driven) or in-situ formation, and 2) damping of misaligned obliquities initially produced by high eccentricity migration (HEM). While multiple lines of empirical evidence support HEM plus damping as the dominant migration mechanism for HJs around AFGK stars \citep{Dawson_2018,Poon_2021}, whether this explanation extends to HJMDs remains an open question. The joint eccentricity--obliquity--\teff{} distribution of HJs \citep{Winn_2015} reveals a broad range of stellar obliquities among planets that orbit hot stars with radiative envelopes located above the Kraft break (i.e. \teff{} $\geq 6100$ K), compared to the higher fraction of aligned obliquities among cooler, low mass planet hosts \citep[c.f. Figure~\ref{fig:spinorbit};][]{Schlaufman_2010,Winn_2010,Rice_2022,Knudstrup_2024}. This finding is broadly consistent with classical tidal theory, which predicts that cool stars, with their comparatively thicker convective envelopes, can efficiently realign large obliquities via strong tidal dissipation compared to earlier-type stars with thinner convective zones \citep{Winn_2010}. This trend has also been hypothesized as a consequence of resonance locking between the planet's orbital frequency and stellar gravity modes \citep[i.e. g-modes;][]{Zanazzi_2024}. Given the emerging population of aligned stellar obliquities among HJMD hosts, can we distinguish between primordial alignment versus a realignment mechanism? While doing so is difficult on a per-system basis, here we consider the prominent theory of tidal obliquity damping and look for hints of this process by considering the obliquity damping timescales of our targets and by estimating the efficiency of KL migration around different types of host stars. Obliquity observations of HJ-hosting, hot subgiants that have cooled to below the Kraft break through post-main sequence evolution show preferentially aligned obliquities; a trend that is consistent with tidal realignment theory \citep{Saunders_2024}.

\subsubsection{Obliquity damping timescales}
Here we calculate each of our target stars' obliquity damping timescales by tides. Explicitly, the obliquity damping timescale is

\begin{eqnarray}
&\tau_{\lambda} = k\, \left( \frac{M_{\mathrm{cz}}}{M_p} \right) \left( \frac{a}{R_\star} \right)^6 \left( \frac{(1-e^2)^{9/2}}{1+3e^2+ (3/8)e^4} \right) \label{eq:taul1} \\ 
&\approx 10.5\, \mathrm{Myr}\, \left( \frac{M_{\mathrm{cz}}}{0.05\, M_\odot} \right)\, \left( \frac{M_p}{M_{\mathrm{Jup}}} \right)^{-1}\, \left( \frac{a}{0.1\, \mathrm{au}} \right)^6 \nonumber \\
&\left( \frac{R_\star}{0.5\, R_\odot} \right)^{-6}\, \left( \frac{(1-e^2)^{9/2}}{1+3e^2+ (3/8)e^4} \right), \label{eq:taul2}
\end{eqnarray}

\noindent where $k$ is a constant, $M_{\mathrm{cz}}$ is the mass of the stellar convective zone, $M_p$ is the planet mass, $a$ is the planet's semi-major axis, $R_\star$ is the stellar radius, and $e$ is the planet's orbital eccentricity \citep{Eggleton_2001}. We follow \cite{Rice_2022} by fixing $k=10^3$ in our calculations, who chose this value as a calibrator in their population-level study to reproduce preferentially aligned in systems younger than 10 Gyrs. To estimate $\tau_\lambda$, we adopt the convective zone mass calculations for low-mass stars at solar metallicity from \cite{Amard_2019}. At field ages of 2-7 Gyrs, $M_{\mathrm{cz}}$ is nearly constant at $\sim 0.08\, M_\odot$ for both of our targets given their similar masses of $\sim 0.5\, M_\odot$. Both HJMDs are consistent with having circular orbits, such that we set $e=0$. Evaluating Eq.~\ref{eq:taul2} for TOI-3714 b and TOI-5293 A b we recover obliquity damping timescales of $7.0^{+1.4}_{-1.2}\times 10^3$ years and $4.1^{+1.2}_{-0.9}\times 10^4$ years, respectively. 

Both system's obliquity damping timescales are shorter than their estimated ages by $\gtrsim 5$ orders of magnitude. As such, we expect that obliquity damping by stellar tides is an efficient process for both HJMDs. Had $\tau_\lambda$ exceeded the likely age of either system, then we could confidently disqualify tidal dissipation as an explanation for the system's aligned obliquity, thus supporting either dynamically cold migration or in-situ formation as the more likely formation pathway. Alas, we are unable to distinguish dynamically cold migration from tidal damping using the $\tau_\lambda$ argument.


\subsubsection{A toy model of the hot Jupiter formation efficiency by Kozai-Lidov migration} \label{sect:toy}
Here we consider the stellar multiplicity fraction of confirmed HJMDs with that of early M dwarfs in the field. Among the 19 HJMDs listed on the NASA Exoplanet Archive, nine are in binary systems. The corresponding stellar multiplicity fraction of $0.47 \pm 0.16$ exceeds the early M dwarf multiplicity fraction in the field \citep[$0.28\pm 0.08$;][]{Winters_2019,Offner_2023}, although the discrepancy is only marginally significant. However, the apparent enhancement in stellar multiplicity fraction among HJMD hosts suggests that KL migration driven by wide binary companions may play an important role in the formation of HJMDs. If true, then the obliquities of early M dwarf HJ hosts should be preferentially misaligned at some point in their evolutionary history, which would leave tidal dissipation as the probable mechanism responsible for producing low HJMD obliquities.

We highlight that this argument is not definitive, such that we seek to corroborate it by developing a toy model of the efficiency of KL migration. We do so by comparing the observed occurrence rates of cold Jupiters to HJs, conditioned on the stellar multiplicity fractions around A, FGK, and early M stars. We note that we ignore mid-to-late M dwarfs as their occurrence rates of hosting hot and cold Jupiters are both consistent with zero \citep{Pass_2023}. 

\begin{figure*}
\centering
\includegraphics[width=0.9\hsize]{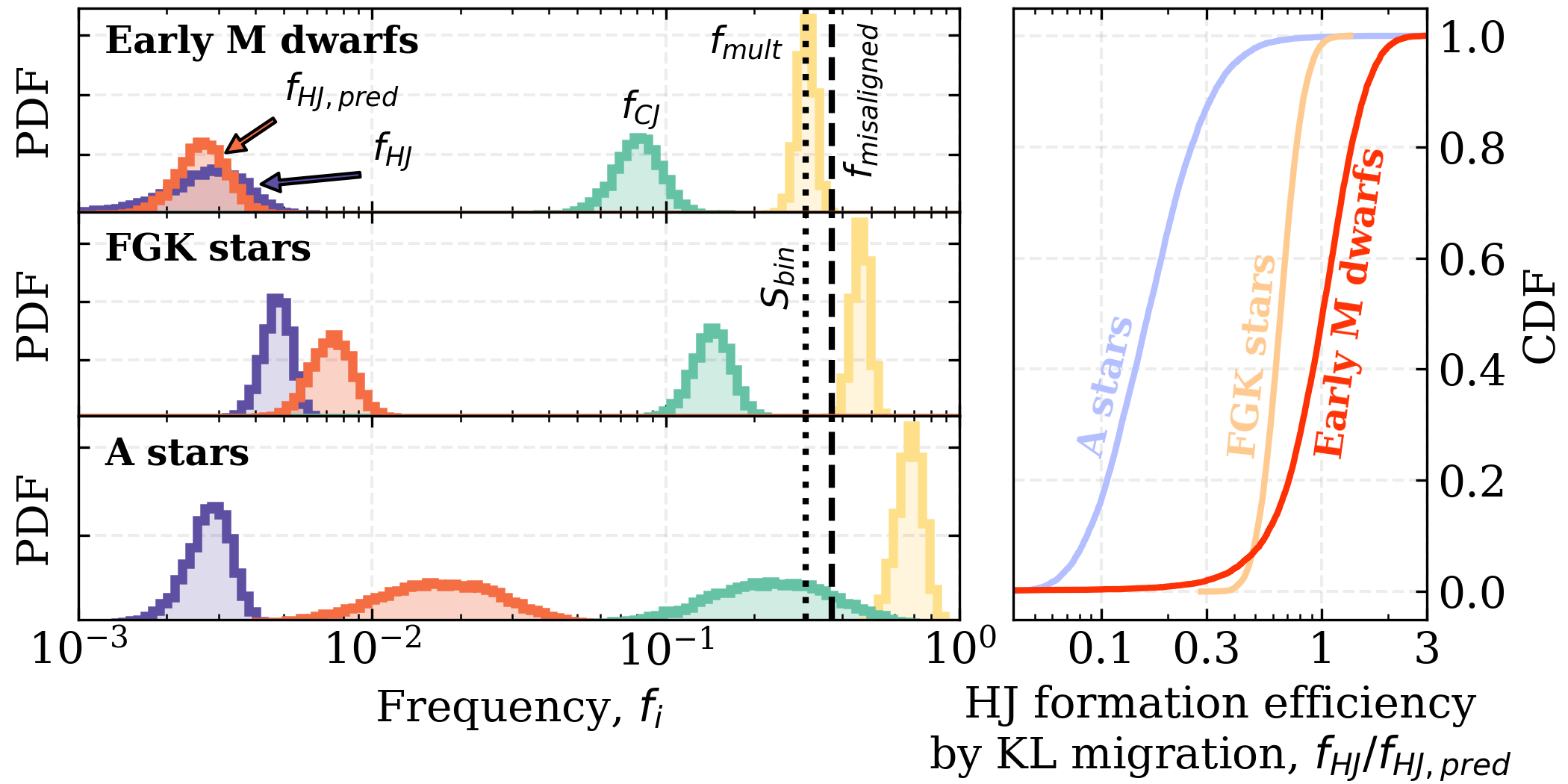}
\caption{Left: A comparison of the various frequency terms featured in Eq.~\ref{eq:hjf} obtained from the literature for early M dwarfs, FGK stars, and A stars. The terms $f_{\mathrm{HJ}}, f_{\mathrm{CJ}}, f_{\mathrm{mult}}$, and $f_{\mathrm{misaligned}}$ denote, respectively, the observed occurrence rates of hot Jupiters, cold Jupiters, the stellar multiplicity fractions, and the fraction of stellar binaries that are misaligned by $\geq 39 \si{\degree}$ drawn from an isotropic distribution (i.e. 36.8\%). The predicted frequency of HJs formed via KL migration is the distribution $f_{\mathrm{HJ,pred}} = f_{\mathrm{CJ}} S_{\mathrm{bin}} f_{\mathrm{mult}} f_{\mathrm{misaligned}}$ (see Eq.~\ref{eq:hjf}). Right: cumulative distribution functions of the HJ formation efficiency by KL migration ($f_{\mathrm{HJ}}/f_{\mathrm{HJ,pred}}$) for each class of star considered herein.}
\label{fig:hjf}
\end{figure*}

\newpage
Our toy model predicts the following occurrence rate of HJs:

\begin{equation}
f_{\mathrm{HJ,pred}} = f_{\mathrm{CJ}} S_{\mathrm{bin}} (f_{\mathrm{mult}} \cdot f_{\mathrm{misaligned}} + f_{\mathrm{pps}}) + f_{\mathrm{in-situ}},
\label{eq:hjf}
\end{equation}

\noindent where $f_{\mathrm{CJ}}$ is the occurrence rate of cold Jupiters, $S_{\mathrm{bin}}$ is the suppression factor of planetary occurrence due to a binary companion compared to around single stars \citep{Moe_2021}, $f_{\mathrm{mult}}$ is the stellar multiplicity fraction, $f_{\mathrm{misaligned}}$ is the fraction of multi-star systems that are sufficiently inclined relative to our line-of-sight to drive KL migration (i.e. $\Delta i = 39.2 \si{\degree}$), and $f_{\mathrm{pps}}$ and $f_{\mathrm{in-situ}}$ are the respective contributions to HJ occurrence from planet-planet scattering and in-situ formation, respectively. Similarly, the product $f_{\mathrm{mult}} f_{\mathrm{misaligned}}$ sets the contribution to HJ occurrence from KL migration. The three terms on the RHS of Eq.~\ref{eq:hjf} represent the source functions of HJs from three prominent formation/migration pathways. Eq.~\ref{fig:hjf} ignores possible contributions to HJ occurrence from alternative pathways such as stellar fly-bys \citep{Shara_2016}. All of these terms are presumed to vary with host spectral type. Caveats to our toy model are presented in Section~\ref{sect:caveats}.

We adopt the following assumptions to evaluate Eq.~\ref{eq:hjf}. Firstly, $f_{\mathrm{misaligned}}=0.368$, which follows from evidence from Gaia DR3 astrometry that the mutual inclination distribution of binary systems with S-type HJs is isotropic \citep{Christian_2025}. Secondly, we set $f_{\mathrm{pps}}=f_{\mathrm{in-situ}}=0$ for convenience as it allows us to evaluate the sole contribution from KL migration to forming the HJ population. However, this is despite evidence for distant planetary companions to HJs that are capable of driving HEM by either KL-like oscillations or planet-planet scattering \citep[e.g.][]{Knutson_2014,Bryan_2016,Zink_2023}. We revisit this assumption in Section~\ref{sect:caveats}. We adopt spectral type-dependent stellar multiplicity fractions, and hot and cold Jupiter occurrence rates from the RV surveys in the literature, which are summarized in Appendix~\ref{app:occ} Table~\ref{tab:occurrence}.

We include the planet suppression factor $S_{\mathrm{bin}}$ in Eq.~\ref{eq:hjf} because there exists strong evidence that planetary occurrence is suppressed by binary companions \citep{Moe_2021}. The physical explanation for this phenomenon is inconclusive but has been hypothesized to be due to lower disk masses in binary systems \citep{Akeson_2019}, altered particle dynamics that inhibit planetesimal and/or pebble growth \citep[e.g.][]{Silsbee_2021}, and dynamical interactions that destabilize planetary orbits \citep{Quarles_2020}. The suppression factor is known to be highly sensitive to the binary's separation \citep{Moe_2021}. Close binaries with $a<1$ au fully suppress S-type planets ($S_{\mathrm{bin}}\sim 0$) while binaries with $a\gtrsim 200$ au exhibit negligible suppression \citep[$S_{\mathrm{bin}}\sim 1$;][]{Moe_2021}. Our toy model calculations adopt stellar multiplicity fractions over all $a$ (see Table~\ref{tab:occurrence}), but these distributions tend to peak between $\sim 10-30$ au for AFGKM stars, where the planet suppression factor is $\approx 30$\%. We account for this effect in our toy model by setting $S_{\mathrm{bin}} = 0.3$. A more detailed calculation that considers $a$-dependent stellar multiplicities and the suppression of cold Jupiters specifically, is beyond the scope of this discussion.


Figure~\ref{fig:hjf} depicts the occurrence rates and stellar multiplicity fractions of early M, FGK, and A-type gas giant hosts along with the HJ frequencies predicted by Eq.~\ref{eq:hjf}. We find that for each spectral type, our toy model of KL migration either over-predicts the frequency of HJs (i.e. $f_{\mathrm{HJ,pred}} > f_{\mathrm{HJ}}$) or finds consistency with the observed value of $f_{\mathrm{HJ}}$. The fact that the predicted frequency of HJs is never less than that observed HJ frequency implies that we do not need to invoke other HJ formation/migration pathways to reproduce $f_{\mathrm{HJ}}$. While this finding does not prove that KL migration is the dominant HJ formation/migration pathway, it does suggest that if KL were solely responsible for producing the observed HJ frequencies, then the process is $\leq 100$\% efficient and the HJ formation efficiency by KL migration ($\equiv f_{\mathrm{HJ}}/f_{\mathrm{HJ,pred}}$) is not equivalent around early M, FGK, and A stars. The right panel of Figure~\ref{fig:hjf} depicts the cumulative distribution functions of the HJ formation efficiency in each spectral type bin. The HJ formation efficiency by KL migration appears to increase monotonically from early to later type stars. Explicitly, we find HJ formation efficiencies by KL migration of $0.16^{+0.12}_{-0.06}$, $0.65^{+0.14}_{-0.12}$, and $1.01^{+0.42}_{-0.36}$, around A, FGK, and early M dwarfs, respectively.

Our toy model suggests that HJ formation by KL migration is more efficient around early M dwarfs than around AFGK stars. Our results also suggest that KL migration may be able to fully account for the population of HJMDs given that the formation efficiency $f_{\mathrm{HJ}}/f_{\mathrm{HJ,pred}}$ around early M dwarfs is consistent with unity. We note that the properties of stellar companions in AFGK binaries are largely incompatible with triggering KL migration of HJs \citep{Ngo_2016} whereas our toy model provides tentative evidence that KL migration by stellar companions may represent the dominant formation channel for HJMDs. Furthermore, because KL migration is expected to produce misaligned stellar obliquities that are preserved in the absence of efficient tidal damping, the emerging population of HJMDs with aligned obliquities (c.f. Figure~\ref{fig:spinorbit}) supports the expectation that M dwarfs, with their comparatively deep convective envelopes, do efficiently damp misaligned obliquities excited by KL oscillations.

\subsubsection{Caveats to our toy model of hot Jupiter formation efficiency by Kozi-Lidov migration} \label{sect:caveats}
Our toy model presented in Section~\ref{sect:toy} sought to estimate the importance of KL migration by comparing the spectral type-dependent occurrence rates of hot and cold Jupiters, conditioned by the fraction of host stars with misaligned binary companions capable of driving KL oscillations. Our simplistic model suffers from a few notable shortcomings.

\begin{enumerate}
\item \emph{It is valid to assume that planet-planet scattering is a negligible contributor to the observed population of HJs around all host star spectral types.} HJs preferentially host distant planetary companions that are capable of driving HEM \citep[e.g.][]{Knutson_2014,Bryan_2016,Zink_2023}. However, it is unclear whether there exists a preferred migration pathway between KL migration and planet-planet scattering such that $f_{\mathrm{pps}}\neq 0$ may be needed to improve the accuracy of our toy model. However, we note that setting $f_{\mathrm{pps}}= 0$ for M dwarf planetary systems is likely reasonable because multiple gas giants are required for planet-planet scattering and the occurrence of cold Jupiters around M dwarfs is relatively low. Conversely, setting $f_{\mathrm{pps}}> 0$ around AFGK stars may reconcile our finding that the HJ formation efficiency around these stars is $<100$\%. There is also evidence for scattering around FGK stars owing to the consistent metallicities between eccentric cold Jupiter hosts and HJ host stars \citep{Dawson_2013}.
\item \emph{It is valid to assume that in-situ formation is a negligible contributor to the observed population of HJs around all host star spectral types.} HJs generally lack nearby planetary companions, which is consistent with expectations from HEM that would destabilize compact, inner planetary systems \citep{Latham_2011,Bryan_2016,Huang_2016}. While counter examples do exist \citep[e.g.][]{Becker_2015}, we posit that $f_{\mathrm{in-situ}}$ is likely a small contributor to Eq.~\ref{eq:hjf} given the plethora of supporting evidence for HEM as the dominant HJ migration mechanism \citep{Dawson_2018,Poon_2021}.
\item \emph{Warm Jupiters are a negligible outcome of inward migrating cold Jupiters and can therefore be ignored on the LHS of Eq.~\ref{eq:hjf}.} There exists evidence from the Kepler mission that warm Jupiters ($P\sim 10-200$ days) preferentially have close planetary companions with low eccentricities and mutual inclinations, thus suggesting that warm Jupiters likely formed in-situ or via smooth disk migration rather than by a HEM channel \citep{Huang_2016}. We posit that warm Jupiters can reasonably be ignored in our toy model.
\item \emph{Inward migrating would-be HJs are not commonly destroyed upon falling onto the central star.} Observational constraints on the rate of HJ ingestion are difficult to obtain because the physical properties of the planetary progenitor are lost after ingestion. However, constraints on the rate of general planetary ingestion is low \citep[$\sim 8$\%;][]{Liu_2024}.  
This is perhaps unsurprising given the large variety of proposed migration stopping mechanisms \citep[e.g.][]{Lin_1996,Kuchner_2002,Rice_2008,Chang_2010,Hasegawa_2011} and justifies our omission of an ingestion term on the LHS of Eq~\ref{eq:hjf}. 
\item \emph{The observed cold Jupiter occurrence rates are representative of cold Jupiter occurrence during the epochs of HJ formation from the primordial cold Jupiter population.} Obtaining observational constraints on whether this assumption is valid is challenging due to low planet detection rates around young, active stars in RV surveys. Although tentative results have suggested a small, but not statistically significant, increase in gas giant occurrence with $P<1000$ days from young ($<400$ Myr) to old FGKM stars \citep{Grandjean_2021,Grandjean_2023}. Establishing the time evolution of cold Jupiter occurrence will require more data from deep surveys of gas giant occurrence around young stars.  
\end{enumerate}

\section{Summary and Conclusions} \label{sect:summary}
We have presented measurements of the RM effect with data from the MAROON-X spectrograph for two hot Jupiters around the early M dwarfs TOI-3714 and TOI-5293 A. Our measurements represent only the second and third detections of the RM effect for hot Jupiters around M dwarfs (HJMDs). 
We summarize our main conclusions below.

\begin{enumerate}
\item We detect RM signatures from both HJMDs at high significance ($7.6-7.8\sigma$). We measure sky-projected stellar obliquities $\lambda=21^{+14}_{-11} \si{\degree}$ and $-12^{+19}_{-14} \si{\degree}$, and projected stellar rotation velocities \vsini{} $=1.06^{+0.15}_{-0.12}$ km/s and $1.13^{+0.17}_{-0.10}$ for TOI-3714 and TOI-5293 A, respectively.
\item We compute deprojected (i.e. 3D) stellar obliquities $\psi= 26^{+11}_{-10} \si{\degree}$ and $24^{+11}_{-10} \si{\degree}$ for TOI-3714 and TOI-5293 A, respectively. Both systems are consistent with being well-aligned.
\item Both HJMDs are in S-type planetary systems with wide binary companions that may be capable of driving Kozai-Lidov (KL) oscillations. Both systems have KL oscillation timescales that are four orders of magnitude shorter than the system ages. The minimum mutual inclination of the planetary and binary orbital planes in the TOI-5293 system is likely capable of driving KL oscillations while it may or may not be sufficient in the TOI-3714 system. A more definitive claim is elusive because only the \emph{minimum} mutual inclination is observationally accessible with Gaia DR3 astrometry.
\item Both systems' obliquity damping timescales are $10^5\times$ shorter than the assumed system ages such that obliquity damping by stellar tides is expected to be an efficient process for both HJMDs.
\item We calculate the HJ formation efficiency by KL migration by comparing the spectral-type dependent hot versus cold Jupiter occurrences rates conditioned on the stellar multiplicity fractions and planet suppression factors by wide binary companions. We conclude that the HJ formation efficiency by KL migration increases with decreasing spectral type and that KL migration can fully account for the occurrence of HJMDs.  
\item If HJMDs preferentially form via high-eccentricity KL migration, then the emerging population of three HJMD hosts with aligned stellar obliquities is consistent with the expectation that M dwarfs do efficiently dampen misaligned obliquities via tidal dissipation.
\end{enumerate}

Stellar obliquity measurements of additional HJMD hosts are needed to confirm that tidal dissipation is driving the emerging population of well-aligned HJMDs. High-resolution optical/near-infrared spectrographs like MAROON-X, as well as similar facilities on $\gtrsim 8$m-class telescopes, can shed light on the distribution of HJMD host obliquities, which will ultimately enable robust comparisons of the dynamical/formation histories of HJMDs to HJs around earlier type stars.

\appendix 

\section{Spectroscopic time series} \label{app:tables}
Our spectroscopic time series measurements presented herein are provided in Tables~\ref{tab:rv_toi3714} and~\ref{tab:rv_toi5293} for TOI-3714 and TOI-5293 A, respectively.

\begin{table}
\caption{MAROON-X spectroscopic time series for TOI-3714. \label{tab:rv_toi3714}}
\centering
\rotatebox{90}{
\begin{tabular}{ccccccccccccccc}
\toprule
Time & RV$_{\mathrm{blue}}$ & $\sigma_{\mathrm{RV,blue}}$ & RV$_{\mathrm{red}}$ & $\sigma_{\mathrm{RV,red}}$ & RV$_{\mathrm{combined}}$ & $\sigma_{\mathrm{RV,combined}}$ & Airmass & BERV & CRX & $\sigma_{\mathrm{CRX}}$ & dLW & $\sigma_{\mathrm{dLW}}$ & H$\alpha$ & $\sigma_{\mathrm{H}\alpha}$ \\
(BJD-2,457,000) & (m/s) & (m/s) & (m/s) & (m/s) & (m/s) & (m/s) &  & (km/s) &  &  &  &  &  &  \\
\midrule
3654.81720 & 32.07 & 9.97 & 48.17 & 6.54 & 43.32 & 5.47 & 1.243 & -2.4293 & 163.60 & 84.66 & 51.61 & 8.61 & 0.715 & 0.031 \\
3654.82864 & 30.80 & 9.89 & 31.46 & 6.48 & 31.26 & 5.42 & 1.199 & -2.4557 & 175.48 & 86.06 & 50.26 & 8.54 & 0.715 & 0.030 \\
3654.84028 & 25.02 & 8.28 & 31.21 & 5.47 & 29.33 & 4.56 & 1.163 & -2.4835 & -8.76 & 75.69 & 20.74 & 7.20 & 0.741 & 0.025 \\
\bottomrule
\multicolumn{15}{l}{Only the first three rows are shown to illustrate the table's contents. The entirety of this table is provided as a csv file in the arXiv source code.} \\
\end{tabular}
}
\end{table}
\begin{table}
\caption{MAROON-X spectroscopic time series for TOI-5293 A. \label{tab:rv_toi5293}}
\centering
\rotatebox{90}{
\begin{tabular}{ccccccccccccccc}
\toprule
Time & RV$_{\mathrm{blue}}$ & $\sigma_{\mathrm{RV,blue}}$ & RV$_{\mathrm{red}}$ & $\sigma_{\mathrm{RV,red}}$ & RV$_{\mathrm{combined}}$ & $\sigma_{\mathrm{RV,combined}}$ & Airmass & BERV & CRX & $\sigma_{\mathrm{CRX}}$ & dLW & $\sigma_{\mathrm{dLW}}$ & H$\alpha$ & $\sigma_{\mathrm{H}\alpha}$ \\
(BJD-2,457,000) & (m/s) & (m/s) & (m/s) & (m/s) & (m/s) & (m/s) &  & (km/s) &  &  &  &  &  &  \\
\midrule
3494.97135 & 45.28 & 8.68 & 46.32 & 5.53 & 46.02 & 4.67 & 2.374 & 28.5248 & 352.22 & 65.92 & -58.43 & 6.88 & 0.535 & 0.095 \\
3494.98254 & 9.27 & 6.84 & 42.19 & 4.45 & 32.41 & 3.73 & 2.080 & 28.5072 & 159.58 & 60.03 & 32.61 & 5.37 & 0.553 & 0.020 \\
3494.99426 & 51.42 & 6.98 & 47.79 & 4.55 & 48.88 & 3.81 & 1.860 & 28.4869 & 89.07 & 45.48 & 10.57 & 5.54 & 0.529 & 0.021 \\
\bottomrule
\multicolumn{15}{l}{Only the first three rows are shown to illustrate the table's contents. The entirety of this table is provided as a csv file in the arXiv source code.} \\
\end{tabular}
}
\end{table}

\section{Toy model of Hot Jupiter formation inputs} \label{app:occ}
Table~\ref{tab:occurrence} summarizes the spectral type-dependent stellar multiplicity fractions, and hot and cold Jupiter occurrence rates from the literature. These values are used in our toy model of HJ formation in Section~\ref{sect:toy}.

\begin{deluxetable*}{lccc}
\tablewidth{0pt}
\tablecaption{Literature values for the frequency of Jovian planets and stellar multiplicity. \label{tab:occurrence}}
\tablehead{
& \colhead{Value} & \colhead{Parameter ranges} & \colhead{Source}}
\startdata
\multicolumn{4}{l}{Hot Jupiter occurrence rates, $f_\mathrm{HJ}$} \\ 
\vspace{-0.2cm} && $P \in [0.8,10]$ days, $R_p \in [0.62,2]R_\mathrm{Jup}$, & \\ \vspace{-0.2cm}
\hspace{1cm} Early M dwarfs & $\mathbf{0.0027\pm 0.0009}$ && 1 \\
&& $M_\star \in [0.45,0.65]M_\odot$ & \\
\hline
\vspace{-0.2cm} && $P \in [0.8,10]$ days, $R_p \in [0.54,1.96]R_\mathrm{Jup}$, & \\
\vspace{-0.2cm}
& $0.0043\pm 0.0005$ && 2 \\
&& all stars in the KIC observed in Q1-Q6 & \\
\vspace{-0.2cm} \hspace{1cm} FGK stars && $P \in [1,10]$ days, $R_p \in [0.71,2.1]R_\mathrm{Jup}$, & \\
\vspace{-0.2cm}
& $0.0057^{+0.0014}_{-0.0012}$ && 3 \\    
&& $T_\mathrm{eff} \in [4700,6500]$ K, $\log{g} \in [3.9,5.0]$ & \\  
& $\mathbf{0.0047\pm 0.0005}$ & - & Weighted average \\
\hline
\vspace{-0.2cm} &&  $P \in [0.9,10]$ days, $R_p \in [0.8,2.5]R_\mathrm{Jup}$, & \\
\vspace{-0.2cm} 
\hspace{1cm} & $0.0026\pm 0.0011$ && 4 \\
&& $M_\star \in [1.4,2.3]M_\odot$ & \\
\vspace{-0.2cm} \hspace{1cm} A stars && $P \in [0.9,10]$ days, $R_p/R_\mathrm{Jup} \in [0.8,2.5]$, & \\
\vspace{-0.2cm}
& $0.0029\pm 0.0005$ && 5 \\ 
&& $M_\star \in [1.4,2.3]M_\odot$ & \\
& $\mathbf{0.0028\pm 0.0005}$ & - & Weighted average \\
\hline
\hline
\multicolumn{4}{l}{Cold Jupiter occurrence rates, $f_{\mathrm{CJ}}$} \\
\vspace{-0.2cm} && $P \in [100,10^4]$ days, $M_p\sin{i} \in [100,10^4]M_\oplus$, & \\ \vspace{-0.2cm}
& $0.080^{+0.037}_{-0.028}$ && 6 \\
&& SpT M, $M_\star \in [0.09,0.75]M_\odot$  & \\
\vspace{-0.2cm} && $a \in [1,5]$ au, $M_p\sin{i} \in [100,6000]M_\oplus$, & \\ \vspace{-0.2cm}
& $0.067^{+0.024}_{-0.019}$ && 7 \\
\hspace{1cm} Early M dwarfs\tablenotemark{a} && $M_\star \in [0.3,0.7]M_\odot$ & \\
\vspace{-0.2cm} && $P \in [100,10^4]$ days, $M_p\sin{i} \in [31.6,3162]M_\oplus$, & \\ \vspace{-0.2cm}
& $0.089^{+0.026}_{-0.022}$ && 8 \\
&& SpT M, $M_\star \in [0.1,0.75]M_\odot$ & \\    
& $\mathbf{0.080^{+0.015}_{-0.014}}$ & - & Weighted average \\    
\hline
\vspace{-0.2cm} && $a \in [1,5]$ au, $M_p\sin{i} \in [100,6000]M_\oplus$, & \\ \vspace{-0.2cm}
\hspace{1cm} FGK stars & $\mathbf{0.143^{+0.022}_{-0.021}}$ && 7 \\
&& $M_\star \in [0.7,1.3]M_\odot$ & \\
\hline
\vspace{-0.2cm} && $a \in [1,5]$ au, $M_p\sin{i} \in [100,6000]M_\oplus$, & \\ \vspace{-0.2cm}
\hspace{1cm} A stars & $\mathbf{0.22^{+0.13}_{-0.09}}$ && 7 \\
&& $M_\star \in [1.3,1.5]M_\odot$ & \\
\hline
\hline
Stellar multiplicity fractions, $f_{\mathrm{mult}}$ &&& \\
\hspace{1cm} Early M dwarfs & $\mathbf{0.30\pm 0.02}$ & $M_\star \in [0.3,0.6]M_\odot$, $\forall\: a$ & 9 \\
\hline
\hspace{1cm} FGK stars & $\mathbf{0.46\pm 0.03}$ & $M_\star \in [0.75,1.25]M_\odot$, $\forall\: a$  & 10 \\
\hline
\hspace{1cm} A stars & $\mathbf{0.68\pm 0.07}$ & $M_\star \in [1.6,2.4]M_\odot$, $\forall\: a$ & 11 \\
\hline
\hline
\multicolumn{4}{l}{Predicted HJ frequency by Kozai-Lidov migration, $f_{\mathrm{HJ,pred}}\equiv f_{\mathrm{CJ}}\cdot S_{\mathrm{bin}}\cdot$\tablenotemark{b} $f_{\mathrm{mult}}\cdot f_{\mathrm{misaligned}}$\tablenotemark{c}} \\
\hspace{1cm} Early M dwarfs & $\mathbf{0.0026^{+0.0018}_{-0.0011}}$ & - & 12 \\
\hspace{1cm} FGK stars & $\mathbf{0.0072^{+0.0018}_{-0.0015}}$ & - & 12 \\
\hspace{1cm} A stars & $\mathbf{0.017^{+0.012}_{-0.007}}$ & - & 12 \\
\hline
\multicolumn{4}{l}{HJ formation efficiency by Kozai-Lidov migration, $f_{\mathrm{HJ}} / f_{\mathrm{HJ,pred}}$} \\
\hspace{1cm} Early M dwarfs & $\mathbf{1.01^{+0.42}_{-0.36}}$ & - & 12 \\
\hspace{1cm} FGK stars & $\mathbf{0.65^{+0.14}_{-0.12}}$ & - & 12 \\
\hspace{1cm} A stars & $\mathbf{0.16^{+0.12}_{-0.06}}$ & - & 12 \\
\enddata
\tablerefs{1) \cite{Gan_2023}, 2) \cite{Fressin_2013}, 3) \cite{Petigura_2018}, 4) \cite{Zhou_2019}, 5) \cite{Beleznay_2022}, 6) \cite{Bonfils_2013}, 7) \cite{Fulton_2021}, 8) \cite{Mignon_2025}, 9) \cite{Winters_2019}, 10) \cite{Raghavan_2010}, 11) \cite{Moe_2021}, 12) this work.}
\tablenotetext{a}{We note that \cite{Bonfils_2013} and \cite{Mignon_2025} include mid-to-late M dwarfs in their respective stellar samples.}
\tablenotetext{b}{$S_{\mathrm{bin}} = 0.3$ is the assumed planet suppression factor by binary companions \citep{Moe_2021}.}
\tablenotetext{c}{$f_{\mathrm{misaligned}} = 0.368$ is the fraction of binary systems that are misaligned by $\geq 39.2^\circ$ based on an isotropic distribution.}
\end{deluxetable*}

\bibliography{refs_master}{}
\bibliographystyle{aasjournalv7}

\begin{acknowledgments}
RC is supported by the Natural Sciences and Engineering Council of Canada (NSERC) through the Discovery Grants program. RC thanks the NOIRLab observing staff at Gemini-North for their assistance with scheduling and executing the observing programs GN-2024A-Q-131 and GN-2024B-Q-132. RC also thanks for following people for their scientific discussions and shared resources that improved the quality of the manuscript: Juliette Becker, Nicole Gromek, Ares Osborn, Bennett Skinner, and Raven Westlake. 

Based on observations obtained at the international Gemini Observatory, a program of NSF NOIRLab, which is managed by the Association of Universities for Research in Astronomy (AURA) under a cooperative agreement with the U.S. National Science Foundation on behalf of the Gemini Observatory partnership: the U.S. National Science Foundation (United States), National Research Council (Canada), Agencia Nacional de Investigaci\'{o}n y Desarrollo (Chile), Ministerio de Ciencia, Tecnolog\'{i}a e Innovaci\'{o}n (Argentina), Minist\'{e}rio da Ci\^{e}ncia, Tecnologia, Inova\c{c}\~{o}es e Comunica\c{c}\~{o}es (Brazil), and Korea Astronomy and Space Science Institute (Republic of Korea).

This work made use of Astropy:\footnote{http://www.astropy.org} a community-developed core Python package and an ecosystem of tools and resources for astronomy \citep{astropyi, astropyii, astropyiii}
 
\end{acknowledgments}


\facilities{Gemini-North MAROON-X).}

\software{\texttt{astropy} \citep{astropyi,astropyii,astropyiii}, \texttt{astroquery} \citep{astroquery_2019}, 
          \texttt{emceee} \citep{foremanmackey13},
          \texttt{lofti\_gaia} \citep{Pearce_2020},
          \texttt{SERVAL} \citep{Zechmeister_2018},
          \texttt{starry} \citep{Luger_2019}
          }

\end{document}